\pdfoutput=1
\documentclass[fleqn,usenatbib,useAMS,onecolumn]{mnras}

\usepackage{graphicx}	
\usepackage{amsmath}	
\usepackage{amssymb}	
\usepackage{multicol}        
\usepackage{bm}		
\usepackage{pdflscape}	
\usepackage{epstopdf}
\usepackage[T1]{fontenc}
\usepackage{ae,aecompl}

\newcommand{\ms}[1]{\textcolor{black}{#1}}

\newcommand{\beq}{\begin{equation}}
\newcommand{\beqa}{\begin{eqnarray}}
\newcommand{\eeq}{\end{equation}}
\newcommand{\eeqa}{\end{eqnarray}}

\newcommand{\simgt}{\lower.5ex\hbox{$\; \buildrel > \over \sim \;$}}
\newcommand{\simlt}{\lower.5ex\hbox{$\; \buildrel < \over \sim \;$}}

\newcommand{\bd}[1]{\mbox{\boldmath $#1$}}

\usepackage[T1]{fontenc}
\usepackage{ae,aecompl}

\usepackage{newtxtext,newtxmath}

\title[Phase-space density of galaxies around clusters]
{
Stacked phase-space density of galaxies around massive clusters: Comparison of dynamical and lensing masses
}

\author[M. Shirasaki et al.]
{Masato Shirasaki$^{1,2}$\thanks{E-mail: masato.shirasaki@nao.ac.jp},
Eiichi Egami$^{3}$,
Nobuhiro Okabe$^{4,5,6}$,
Satoshi Miyazaki$^{1}$
\\
$^{1}$National Astronomical Observatory of Japan, 
Mitaka, Tokyo 181-8588, Japan\\
$^{2}$The Institute of Statistical Mathematics, Tachikawa, Tokyo 190-8562, Japan\\
$^{3}$Steward  Observatory,  University  of Arizona, 
933 North Cherry Avenue, Tucson, AZ 85721, USA\\
$^{4}$Physics Program, Graduate School of Advanced Science and Engineering, Hiroshima University, 
Hiroshima 739-8526, Japan\\
$^{5}$Hiroshima Astrophysical Science Center, Hiroshima University, 
Hiroshima 739-8526, Japan\\
$^{6}$Core Research for Energetic Universe, Hiroshima University, 
Hiroshima 739-8526, Japan
}
\begin{document}

\date{}

\pagerange{\pageref{firstpage}--\pageref{lastpage}} \pubyear{2021}

\maketitle

\label{firstpage}

\begin{abstract}
We present a measurement of average histograms of 
line-of-sight velocities over pairs of galaxies and galaxy clusters.
Since the histogram can be measured at different galaxy-cluster separations, this observable is commonly referred to as 
the stacked phase-space density.
We formulate the stacked phase-space density based on a halo-model approach 
so that the model can be applied to real samples of 
galaxies and clusters.
We examine our model by using an actual sample of massive clusters with known weak-lensing masses and spectroscopic observations of galaxies around the clusters.
A likelihood analysis with our model enables us to infer
the spherical-symmetric velocity dispersion of observed galaxies 
in massive clusters.
We find the velocity dispersion of galaxies surrounding clusters with their lensing masses of $1.1\times10^{15}\, h^{-1}M_{\odot}$ 
to be $1180^{+83}_{-70}\, \mathrm{km/s}$ at the 68\% confidence level.
Our constraint confirms that the relation between the galaxy velocity dispersion and the host cluster mass in our sample is consistent with the prediction in dark-matter-only N-body simulations under General Relativity.
Assuming that the Poisson equation in clusters can be altered by an effective gravitational constant of $G_\mathrm{eff}$, 
our measurement of the velocity dispersion can place a tight constraint of 
\ms{$0.88 < G_\mathrm{eff}/G_\mathrm{N} < 1.29\, (68\%)$} at length scales of a few Mpc
about $2.5$ Giga years ago, where $G_\mathrm{N}$ is the Newton's constant.
\end{abstract}


\section{INTRODUCTION}

Galaxy clusters are 
the largest gravitationally bound objects in the universe.
The abundance of clusters at different redshifts 
as a function of their masses is known to be a powerful probe of gravitational growth in cosmic mass density 
as well as expansion history of our universe \citep[e.g.][for a review]{2011ARA&A..49..409A}.
Although previous astronomical surveys at different wavelengths 
have enabled us to construct a large sample of galaxy clusters,
the constraining power of cosmological analyses with clusters is 
currently limited by the uncertainty in estimates of cluster masses \citep[e.g.][]{2009ApJ...692.1060V, 2010ApJ...708..645R, 2014MNRAS.440.2077M, 2016A&A...594A..24P, 2016ApJ...832...95D, 2019ApJ...878...55B, 2020PhRvD.102b3509A}.
Ongoing and upcoming surveys will further increase the number of galaxy clusters \citep[e.g.][]{2012arXiv1209.3114M, 2016arXiv161002743A}, allowing us to measure the cluster abundance with a high precision.
Hence, an accurate estimate of individual cluster masses is
an important and urgent task in studies of galaxy clusters.

Since the first application to the Coma cluster \citep{1937ApJ....86..217Z}, 
kinematics of galaxies in a cluster has been commonly 
used to infer the cluster mass so far.
We refer to the mass estimate by galaxy kinematics as 
the dynamical mass.
The underlying assumption in the dynamical mass estimate
is that the motions of galaxies inside a cluster is 
a good tracer of the gravitational potential in the system.
A set of cosmological N-body simulations has shown that the velocity dispersion of N-body particles inside cluster-sized halos exhibits a tight correlation to the true halo masses with a small scatter, regardless of assumed cosmological models \citep{2008ApJ...672..122E}.
In practice, there are several complicating factors to use the kinematic information of member galaxies as a proxy of the cluster mass.
\citet{2006A&A...456...23B} have studied possible biases of the dynamical mass estimate based on the velocity dispersion of member galaxies with a hydrodynamical simulation.
They found that the degree in the mass bias depends 
on the number of member galaxies used.
High-resolution zoom-in simulations of galaxy clusters
have revealed that the scaling relation between the velocity dispersion of galaxies
and the true cluster mass can differ from the dark-matter counterpart, and that 
the difference depends on the selection
of member galaxies \citep{2010ApJ...708.1419L, 2013MNRAS.430.2638M, 2018MNRAS.474.3746A}.
Besides the bias, \citet{2013ApJ...772...47S} have shown that projection effects of galaxies selected by spectroscopic observations can make the scatter in the scaling relation larger.
\citet{2010MNRAS.408.1818W} also found that the velocity dispersion in the line-of-sight direction can be correlated with the orientation of large-scale structures around clusters.

Apart from the dynamical mass estimate on an individual basis, 
galaxy kinematics in high-density environments provides 
a means of testing the theory of gravity.
Numerical simulations have shown that velocity statistics of galaxies around clusters have a great potential to test gravity on non-linear scales \citep{2010PhRvD..81j3002S, 2012PhRvL.109e1301L, 2014PhRvL.112v1102H, 2014MNRAS.445.1885Z}.
For a comprehensive search of modifications of gravity, 
it is essential to investigate the kinematics of galaxies outside the virial region of clusters, because a wide class of modified gravity theories can reduce to General Relativity (GR) in high-density regions \citep[e.g.][for a review]{2019arXiv190803430B}.
However, galaxy kinematics in the outskirts of clusters is
subject to projection effects on a cluster-by-cluster basis.
\citet{2011Natur.477..567W} have demonstrated that a stacking analysis of the line-of-sight velocity distribution of galaxies around optically-selected clusters can be used to infer the gravitational redshift of light at scales beyond the cluster virial regime.
\citet{2012PhRvL.109e1301L} have also proposed the velocity dispersion in the stacked distribution at different galaxy-cluster separations for a robust test of gravity on scales of 1-30 Mpc.
The stacked velocity distribution of galaxies around clusters, referred to as the stacked phase-space density, has drawn much attention as a probe of the average dynamical mass estimate and galactic orbits in the cluster regime \citep{2009MNRAS.399..812W}, the physics of star formation quenching \citep{2015ApJ...806..101H, 2019ApJ...878....9A}, and large-scale galaxy infall \citep{2021MNRAS.502.1041A, 2020MNRAS.499.1291T}.

The stacking analysis of galaxies around clusters in phase space\footnote{Note that the caustic method \citep{1997ApJ...481..633D} is closely related to our stacking analysis in phase space.
Our theoretical framework can be regarded as natural extensions of the caustic method.
The caustic method measures the velocity dispersion of galaxies as a function of cluster-centric radii $R$, while our model predicts the distribution of galaxy velocity as a function of $R$.}
contains rich information about structure formation and possible modifications of gravity, 
whereas its theoretical model is still developing.
\citet{2009MNRAS.399..812W} have developed a theoretical model of the phase-space density of galaxies inside a cluster by assuming a spherically symmetric system in dynamical equilibrium, while \citet{2013MNRAS.429.3079M} have proposed a fast, efficient method to predict the phase-space density for a given three-dimensional velocity distribution of galaxies in a cluster.
\citet{2013MNRAS.431.3319Z} have studied the pairwise velocity distribution for galaxy-cluster pairs by using a semi-analytic galaxy model, and have built a phenomenological model of the velocity distribution as a function of galaxy-cluster separation lengths.
On the other hand, \citet{2013PhRvD..88b3012L} developed a semi-analytic approach based on the halo model \citep{2002PhR...372....1C} and the perturbation theory of large-scale structures. 
Recently, \citet{2019MNRAS.489.1344H} have proposed a two-component model of the three-dimensional phase space distribution of halos surrounding clusters up to $50\, h^{-1}\mathrm{Mpc}$ from cluster centres based on N-body simulations.

In this paper, we aim at constraining the velocity dispersion of galaxies within clusters with a measurement of the stacked phase-space density and lensing masses of clusters.
For this purpose, we adopt the halo-model prescription
as in \citet{2013PhRvD..88b3012L}, while we improve the previous model so that it can be applied 
to realistic observational data.
In our model, we can include 
a wide distribution of cluster masses 
(not restricted by mass-limited cases), 
contributions from satellite galaxies in single clusters,
and large-scale pairwise velocity statistics between two halos.
We then examine our model with a set of
spectroscopic observations of galaxies around low-redshift massive clusters.
For the sample of massive clusters, 
we use 23 clusters at $0.15 \le z \le 0.30$ from the Local Cluster Substructure Survey (LoCuSS\footnote{\url{http://www.sr.bham.ac.uk/locuss/}}).
For the galaxy sample around the LoCuSS clusters, we use
a highly-complete spectroscopic observation by
the Arizona Cluster Redshift Survey (ACReS\footnote{\url{https://herschel.as.arizona.edu/acres/acres.html}}). 
Making the best use of precise weak-lensing mass 
estimates of the LoCuSS clusters \citep{2016MNRAS.461.3794O}, we specify key ingredients in our model such as the cluster mass distribution and the spatial distribution of the ACReS galaxies around the clusters.
Through a likelihood analysis, we infer kinematic information about the ACReS galaxies and compare the simulation-calibrated prediction \citep{2008ApJ...672..122E} under a standard $\Lambda$ cold dark matter ($\Lambda$CDM) cosmology.
In the end, we study a possible deviation of GR with our measurement of the velocity dispersion of galaxies with known lensing cluster masses.

The paper is organised as follows. 
In Section~\ref{sec:data}, we describe the observational data of 
galaxies and clusters used in this paper.
We also provide a short summary of N-body simulation data to assess statistical errors of our measurement and calibrate our analytic model in Section~\ref{sec:data}.
In Section~\ref{sec:stacked_density},
we present an overview of the stacking analysis in phase space,
introduce our halo-based model, and set model parameters by using the observational data.
The results are presented in Section~\ref{sec:results}.
Finally, the conclusions and discussions are provided in Section~\ref{sec:conclusion}.
Throughout this paper, we assume the cosmological parameters which are consistent with the observation of cosmic microwave backgrounds by the Planck satellite \citep{2016A&A...594A..13P}.
To be specific, we adopt
the cosmic mass density $\Omega_{\mathrm{m0}}= 0.31$,
the baryon density $\Omega_{\mathrm{b0}}= 0.048$,
the cosmological constant $\Omega_{\Lambda}=1-\Omega_{\mathrm{m0}} = 0.69$,
the present-day Hubble parameter $H_0 = 100h\, \mathrm{km}/\mathrm{s}/\mathrm{Mpc}$ with $h= 0.68$,
the spectral index of primordial curvature perturbations 
$n_s= 0.96$, and 
the linear mass variance smoothed over $8\, h^{-1}\mathrm{Mpc}$ $\sigma_8= 0.83$.
We also refer to $\log$ as the logarithm with base 10, while $\ln$ represents the natural logarithm.

\section{DATA}\label{sec:data}

In this section, we describe our observational data set to study the phase-space density of galaxies around massive clusters.
Table~\ref{tab:cluster} summarises the basic property of the cluster sample used in this paper.
In addition, we use a public N-body simulation data to estimate our model uncertainties due to the large-scale pairwise velocity between dark matter halos and statistical errors in our measurement of the phase-space density.

\begin{table*}
\caption{
    The cluster sample. Column (1) Cluster name; col.~(2) Mean redshift of cluster members; col.~(3) Total number of unique target spectra; col.~(4) Secure redshifts; col.(5) Cluster mass in $10^{14}\, h^{-1} M_{\odot}$ \citep{2016MNRAS.461.3794O}. The mass $M_{\mathrm{200b}}$ is defined by the spherical overdensity mass with respect to 200 times the mean mass density in the universe. \label{tab:cluster}
	}
\begin{tabular}{@{}ccccc}
\hline
Cluster & $\langle z \rangle$ & Number of unique & Secure redshits & $M_{\mathrm{200b}}$  \\ 
name & & target spectra & & $(10^{14}\, h^{-1}M_{\odot})$ \\ 
\hline
Abell68 & 0.2546 & 865 & 731 & $8.73^{+2.02}_{-1.66}$ \\
Abell115 & 0.1971 & 710 & 414 & $11.55^{+6.12}_{-3.69}$ \\
Abell209 & 0.206 & 676	& 608 & $17.71^{+3.74}_{-2.98}$ \\
Abell267 & 0.23	& 504 & 290 & $8.37^{+1.87}_{-1.55}$ \\
Abell291 & 0.196 & 529 & 432 & $9.02^{+3.55}_{-2.40}$ \\
Abell383 & 0.1883 & 1260 & 958 & $7.17^{+2.20}_{-1.67}$ \\
Abell586 & 0.171 & 777 & 713 & $8.63^{+3.53}_{-2.36}$ \\
Abell611 & 0.288 & 1030 & 701 & $12.25^{+2.81}_{-2.39}$ \\
Abell697 & 0.282 & 1038	& 859 & $14.84^{+6.21}_{-3.83}$ \\
Abell963 & 0.205 & 1298	& 1137 & $9.84^{+2.22}_{-1.82}$ \\
Abell1689 & 0.1832 & 1301 & 1071 & $13.63^{+2.34}_{-1.99}$ \\
Abell1758 & 0.28 & 1523 & 1273 & $7.51^{+2.45}_{-1.86}$ \\
Abell1763 & 0.2279 & 1005 & 836& $23.83^{+6.00}_{-4.40}$ \\
Abell1835 & 0.2528 & 1256 & 1008 & $12.73^{+2.78}_{-2.32}$ \\
Abell1914 & 0.1712 & 945 & 781 & $13.02^{+3.59}_{-2.70}$ \\
Abell2219 & 0.2281 & 725 & 571 & $15.81^{+4.60}_{-3.23}$ \\
Abell2390 & 0.2329 & 1041 & 822 & $14.30^{+2.94}_{-2.45}$ \\
Abell2485 & 0.2472 & 1053 & 682 & $7.86^{+2.30}_{-2.45}$ \\
RXJ1720.1+2638 (R1720) & 0.164 & 1181 & 1019 & $7.52^{+3.49}_{-2.30}$ \\
RXJ2129.6+0005 (R2129) & 0.235 & 996 & 895 & $7.69^{+4.20}_{-2.54}$ \\
ZwCl0104.4+0048 (Z348) & 0.254 & 924 & 786 & $3.10^{+2.22}_{-1.28}$ \\
ZwCl0857.9+2107 (Z2089) & 0.2347 & 1006	& 571 & $3.67^{+2.00}_{-1.42}$ \\
ZwCl1454.8+2233 (Z7160) & 0.2578 & 1183	& 774 & $6.55^{+6.10}_{-2.75}$ \\
\hline
\end{tabular}
\end{table*}

\subsection{LoCuSS}

LoCuSS is a multi-wavelength survey of X-ray luminous clusters 
at $0.15 \le z \le 0.30$ drawn from the ROSAT All Sky Survey cluster catalogues \citep{1998MNRAS.301..881E, 2000MNRAS.318..333E, 2004A&A...425..367B}. The LoCuSS cluster sample consists of 50 clusters and its selection criteria is given by
(1) $-25^{\circ} < \mathrm{Dec} < +65^{\circ}$ (2) the interstellar column density $n_{\mathrm{H}} \le 7\times 10^{20}\, \mathrm{cm}^2$ (3) $0.15 \le z \le 0.30$ (4) $L_{\mathrm{X}}/E(z) > 4.1\times10^{44}\, \mathrm{erg}\mathrm{s}^{-1}$ where $L_{\mathrm{X}}$ is an X-ray luminosity in the 0.1-2.4 keV band
and $E(z)=\sqrt{\Omega_{\mathrm{m0}}(1+z)^3+\Omega_{\Lambda}}$
with $\Omega_{\mathrm{m0}}=0.3$ and $\Omega_{\Lambda}=0.7$.
Full details of the selection function are available in \citet{2016MNRAS.456L..74S}.
The sample is therefore purely X-ray luminosity limited 
and \citet{2010PASJ...62..811O} shows that the sample is statistically indistinguishable from a volume-limited sample.
The great advantage in the LoCuSS cluster sample is that the precise estimates of individual weak-lensing masses are available \citep{2010PASJ...62..811O, 2016MNRAS.461.3794O}.
For the mass estimates, we adopt the latest results in \citet{2016MNRAS.461.3794O},
which performed the measurement of weak lensing mass for the LoCuSS clusters with possible systematic biases of a $<4\%$ level.
To be specific, we use the mass estimates based on a fitting
of the tangential shear profiles by the Navarro-Frenk-White (NFW) profile \citep{1996ApJ...462..563N, 1997ApJ...490..493N}.
The fitting has been performed with corrections of the shear calibration and the contamination of member galaxies (see the fitted results for Table~B1 in \citet{2016MNRAS.461.3794O}).
Throughout this paper, we define the cluster mass by a spherical overdensity mass with respect to 200 times the mean mass density in the universe, i.e. $M_{\mathrm{200b}} = 4\pi/3 \times 200 \times \bar{\rho}_{\mathrm{m0}} \, r^3_{\mathrm{200b}}$ where
$\bar{\rho}_{\mathrm{m0}}$ is the present-day mean mass density and $r_{\mathrm{200b}}$ is the halo radius in the unit of comoving distance.

\subsection{ACReS}

ACReS is an optical spectroscopic survey programme to observe 30 clusters among the LoCuSS sample with MMT/Hectospec \citep[e.g.,][]{2015ApJ...806..101H}.
These 30 clusters were selected from the parent 
LoCuSS sample on the basis of being accessible to the Subaru telescope
on the nights allocated to ACReS \citep{2010PASJ...62..811O}.
Hence, the cluster selection is not subject to any biases due to
the dynamical state of individual cluster.
Target galaxies around the clusters are primary selected 
by the K-band magnitude of $M^{*}_{\mathrm{K}}(z_{\mathrm{cl}})+1.5$ or brighter,
where $z_{\mathrm{cl}}$ is the cluster redshift
and $M^{*}_{\mathrm{K}}$ is the K-band magnitude corresponding 
to a characteristic galaxy luminosity \citep{1976ApJ...203..297S}.
The selection in ACReS aims at producing an approximately stellar mass-limited sample down to $M_{*} \sim 2 \times 10^{10}\, M_{\odot}$. 
Higher priorities are given to target galaxies also detected 
at 24$\mu$m to obtain a virtually complete census of 
obscured star formation in the cluster population.
For the data set used in this paper, the spectroscopic observations
have been performed across a fixed $52^{\prime} \times 52^{\prime}$ field-of-view around each cluster,
which corresponds to the field-of-view of the UKIRT/WFCam near-infrared imaging data used for
the ACReS target selection.
Details of the target selection are given 
by \citet{2013ApJ...775..126H}. 
Eleven of our 30 clusters were also observed by the Hectospec Cluster Survey \citep[HeCS][]{2013ApJ...767...15R}, providing redshifts for additional 971 cluster members.
Redshifts for further 112, 92 and 49 members of clusters RXJ1720.1+2638, Abell 383 and Abell 209 are included from
\citet{2011ApJ...741..122O}, \citet{2014ApJ...783...52G} and \citet{2003A&A...397..431M}, respectively.
Because the additional member galaxies are selected based on optical imaging data,
they are not necessarily stellar mass-limited. 
Nevertheless, we expect that these galaxies do not affect our results significantly, because they account for only $\sim6$\% of galaxies in the analysis and our theoretical model does not use the information about stellar masses in a direct manner.
In this paper, we only use 23 LoCuSS clusters with 
the weak-lensing mass inferred by \cite{2016MNRAS.461.3794O}.
For our analysis, we only use those redshifts that were classified as secure in the ACReS catalog, meaning that the measurements were based on the detection of 
multiple high-significance spectral features.
For individual clusters, the numbers of target spectra and galaxies with secure redshift estimates are shown in Table~\ref{tab:cluster}.
We use 16701 galaxies around 23 clusters in the stacked analysis.

\subsection{hCOSMOS}\label{subsec:hCOSMOS}
To quantify environmental effects around massive clusters,
we use the spectroscopic data sets taken in the hCOSMOS survey
\citep{2018ApJS..234...21D}.
The hCOSMOS is the redshift survey of the COSMOS field conducted with the Hectospec spectrograph on the MMT.
In the central 1 $\mathrm{deg}^2$ of the original COSMOS field, 
the survey allows us to study $>90\%$ of galaxies with a limiting magnitude of 20.6 in the $r$ band. 
The hCOSMOS survey also includes 1701 new redshifts 
in the COSMOS field.
To mimic an observation of field galaxies in ACReS, we impose the selection in the hCOSMOS data sets by the stellar mass of each galaxy being $M_{*} \ge 2\times 10^{10}\, M_{\odot}$ over redshifts.
This selection leaves 1061 galaxies available in our analysis.
We then define the field galaxies around a cluster by 
randomly setting an angular position in the COSMOS field
and finding the hCOSMOS galaxies around the randomly-selected pseudo-cluster position 
across a $52^{\prime} \times 52^{\prime}$ field-of-view.
This random process will be repeated as much as needed for statistical analyses.
We find that 10,000 random sampling for each cluster is sufficient to obtain the converged result for the estimate of statistical errors in the stacked phase-space density of galaxies around the clusters in ACReS.

\subsection{$\nu^2$GC simulation}\label{subsec:nu2gc}
The hCOSMOS data is used to evaluate effects of field galaxies in our analysis, while it does not contain the information of cluster members.
To evaluate realistic statistical errors in our analysis, 
we need to populate a mock cluster in the hCOSMOS data.
For this purpose, we use a publicly available halo catalogue at $z=0.19$ 
provided by the $\nu^2$GC collaboration\footnote{The data are available at \url{https://hpc.imit.chiba-u.jp/~nngc/}.}.
The halo catalogue has been constructed with the largest-volume run called $\nu^2$GC-L run, which consists of $8192^3$ dark matter particles in a box of $1.12\, h^{-1}\mathrm{Gpc}$ \citep[see][for details of the simulations]{2015PASJ...67...61I}.
In the simulations, the following cosmological parameters were adopted: $\Omega_{\mathrm{m0}}= 0.31$,
$\Omega_{\mathrm{b0}}= 0.048$,
$\Omega_{\Lambda}=1-\Omega_{\mathrm{m0}} = 0.69$,
$h= 0.68$,
$n_s= 0.96$, and $\sigma_8= 0.83$.
These are consistent with Planck \citep{2016A&A...594A..13P}.
We work with the halo catalogue produced with the ROCKSTAR halo finder \citep{2013ApJ...762..109B} in this paper.

We summarise how to evaluate the statistical error in our measurement of the phase-space density by combining the hCOSMOS data and $\nu^2$GC halo catalogue in Section~\ref{subsubsec:measurement}.
We also use the $\nu^2$GC halo catalogue to study the pairwise velocity between two distinct dark matter halos on large scales (see Appendix~\ref{apdx:two_halo_sim}).


\section{STACKED PHASE-SPACE DENSITY}\label{sec:stacked_density}

In this section, we summarise basics in the analysis of stacked phase-space density of galaxies around clusters. We also describe a theoretical framework to predict the stacked phase-space density
within the halo-model approach \citep{2002PhR...372....1C}.
Similar theoretical models have been found in the literature \citep[e.g.][]{2004MNRAS.352.1302V,2009MNRAS.392..917M,2019MNRAS.488.4984V}. 

\subsection{Basics}

Let us assume that we have a sample of galaxy clusters with secure redshift measurements. Then consider a case that we perform a spectroscopic observation of galaxies around individual clusters.
Using the sample of clusters and galaxies, 
we can produce the histogram of the pairwise velocity 
between clusters and galaxies. We denote the histogram as 
${\cal H}(\hat{v})$, where $\hat{v}$ is the estimator of the pairwise velocity for a cluster-galaxy pair. 
The pairwise velocity ${\hat v}$ is estimated in the rest frame of individual clusters:
\beqa
\hat{v} \equiv c \frac{z_{\mathrm{g}}-z_{\mathrm{cl}}}{1+z_{\mathrm{cl}}},
\eeqa
where $c$ is the speed of light, $z_{\mathrm{g}}$ and $z_{\mathrm{cl}}$ are the redshifts of galaxy and cluster, respectively.
In an expanding universe, one can find
\beqa
\hat{v} = \frac{H(z_{\mathrm{cl}})}{1+z_{\mathrm{cl}}}(\bd{r}\cdot\hat{\bd{n}}) + \bd{v}_{\mathrm{gc}}\cdot\hat{\bd{n}},
\eeqa
where $H(z)$ is the Hubble parameter at $z$,
$\hat{\bd{n}}$ is the unit vector pointing to 
a line-of-sight direction, $\bd{r}$ is the comoving separation distance from the cluster to the galaxy, 
and $\bd{v}_{\mathrm{gc}}$ represents the relative velocity of 
the cluster-galaxy pair.
Throughout this paper, we assume that the histogram 
${\cal H}(\hat{v})$ can be measured at different $r_{p}$, 
where $r_{p}$ is the comoving separation between the cluster-galaxy pair in a direction perpendicular to the line of sight.

Since the histogram is constructed from the number count of galaxy-cluster pairs as a function of $\hat{v}$ and $r_{p}$, it is formally written as
\beqa
{\cal H}(\hat{v}\, |\, r_{p}) &=& {\cal H}^{-1}_0 (r_p)\, \left\{
\int \mathrm{d}r_{\parallel}\, 2 \pi r_p \, 
{\cal F}_{\mathrm{gc}}\left(\bd{r} = (\bd{r}_{p}, r_{\parallel}) \, \Bigg|\,  \hat{v}-\frac{H(z_{\mathrm{cl}})}{1+z_{\mathrm{cl}}}r_{\parallel} \right) + {\cal H}_{\mathrm{int}}(r_p) \right\}, \label{eq:hist}\\
{\cal F}_{\mathrm{gc}}(\bd{r} \, | \, v_{\mathrm{gc},\parallel}) &=& \bar{n}_{\mathrm{g}}\bar{n}_{\mathrm{cl}}
\left[1+\xi_{\mathrm{gc}}(r)\right]P_{\mathrm{gc}}(\bd{r}, v_{\mathrm{gc},\parallel}), \label{eq:pairs_phase_space}
\eeqa
where $r_{\parallel} = \bd{r}\cdot\hat{\bd{n}}$, 
$v_{\mathrm{gc},\parallel} = \bd{v}_{\mathrm{gc}}\cdot\hat{\bd{n}}$,
$\xi_{\mathrm{gc}}$ is the galaxy-cluster correlation function in real space, $P_{\mathrm{gc}}$ is the probability distribution function of the line-of-sight pairwise velocity, and
$\bar{n}_{\mathrm{g}}$ and $\bar{n}_{\mathrm{cl}}$ are the mean number density of galaxies and clusters, respectively.
In Eq.~(\ref{eq:hist}), the term ${\cal H}_{\mathrm{int}}$ represents the contribution from uncorrelated large-scale structures and we assume ${\cal H}_{\mathrm{int}}$ to be a constant number for a given $r_p$.
The normalisation ${\cal H}_{0}$ is set by 
$\int_{v_{\mathrm{min}}}^{v_{\mathrm{max}}} \mathrm{d}\hat{v} \, {\cal H}(\hat{v})=1$.
In this paper, we set $v_{\mathrm{min}}=-5000\, \mathrm{km}\,\mathrm{s}^{-1}$ and $v_{\mathrm{max}}=5000\, \mathrm{km}\,\mathrm{s}^{-1}$.
Note that we ignore possible selection effects of galaxies as a function of $\bd{r}$ in Eq.~(\ref{eq:hist}). We expect that our results can be less affected by the selection effects, as long as the histogram ${\cal H}(\hat{v})$ is normalised at different $r_{p}$.

\subsection{A halo model}
\label{subsec:halo_model}

We here summarise a theoretical model of ${\cal H}(\hat{v})$ 
based on a halo-based approach.
In the standard halo model \citep[e.g.][for a review]{2002PhR...372....1C}, the phase-space density of clusters can be expressed as
\beqa
f_{\mathrm{cl}}(\bd{x}, v_{\parallel}) = \sum_{i} S(M_{i}) \delta^{(3)}_{\mathrm{D}}(\bd{x}-\bd{x}_{i})
\delta^{(1)}_{\mathrm{D}}(v_{\parallel}-v_{\parallel, i}), \label{eq:f_cl}
\eeqa
where the index $i$ runs over all dark matter halos in the universe at a given redshift, 
$\delta^{(n)}_{\mathrm{D}}(\bd{x})$ is the Dirac delta function 
in $n$-dimensional space, 
$S(M)$ represents the selection function of clusters, $\bd{x}_{i}$ and $v_{\parallel, i}$ are the position and the line-of-sight velocity of $i$-th dark matter halo, respectively.
We here assume that the selection of clusters can depend on the halo mass alone for simplicity.
Similarly, one can express the phase-space density of galaxies
by using a halo occupation distribution (HOD):
\beqa
f_{\mathrm{g}}(\bd{x}, v_{\parallel}) = \sum_{i} N(M_{i}) u_{\mathrm{g}}(\bd{x}-\bd{x}_{i}\, |\,M)
w_{\mathrm{g}}(v_{\parallel}-v_{\parallel, i}\, |\, \bd{x}-\bd{x}_{i}, M), \label{eq:f_g}
\eeqa
where $N(M)$ is the number of galaxies in a halo of $M$,
$u_{\mathrm{g}}(\bd{x}|M)$ is the number density profile of galaxies,
$w_{\mathrm{g}}(v|\bd{r},M)$ is the velocity distribution function of galaxy at the halo-centric radius of $\bd{r}$ in the rest frame of dark matter halos.
Note that we set
$\int \mathrm{d}\bd{x}\, u_{g}(\bd{x}|M) = 
\int \mathrm{d}v\, w_{g}(v|\bd{r},M) = 1$.
Also, we assume that the number of galaxies in single halos is determined by the halo mass alone.
In the following, we omit the $M$-dependence of $u_{\mathrm{g}}$ and $w_{\mathrm{g}}$
for sake of simplicity.

One can express the number density of cluster-galaxy pairs in phase space as
\beqa
{\cal F}_{\mathrm{gc}}(\bd{r}\, |\, v)
= \int \mathrm{d}v_2\, \langle f_{\mathrm{g}}(\bd{x}_1, v_1) f_{\mathrm{cl}}(\bd{x}_2, v_2)\rangle, \label{eq:pairs_phase_space_def}
\eeqa
where $\bd{r}=\bd{x}_{1}-\bd{x}_2$,
$v = v_1-v_2$ and
$\langle \cdots \rangle$ represents an ensemble average.
Note that we omit the index of $\parallel$ in the line-of-sight velocity in the rest of this section.
In Eq.~(\ref{eq:pairs_phase_space_def}), we introduce a marginalisation over the cluster velocity $v_{2}$ because we are interested in the phase-space density as a function of the relative velocity $v$.
Using Eqs.~(\ref{eq:f_cl}) and (\ref{eq:f_g}), one can find
\beqa
\langle f_{\mathrm{g}}(\bd{x}_1, v_1) f_{\mathrm{cl}}(\bd{x}_2, v_2)\rangle &=& 
\Big\langle 
\int \mathrm{d}M \, \mathrm{d}M'\, 
\mathrm{d}\bd{y}_1\,  \mathrm{d}\bd{y}_2\,
\mathrm{d}p_1\,  \mathrm{d}p_2\, 
\sum_{i,j} N(M) \, S(M')\, 
u_{\mathrm{g}}(\bd{x}_1-\bd{y}_1)\,
\delta^{(3)}_{\mathrm{D}}(\bd{x}_2-\bd{y}_2)\,
w_{\mathrm{g}}(v_1-p_1\, |\, \bd{x}_1-\bd{y}_1)
\nonumber \\
&\times&
\delta^{(1)}_{\mathrm{D}}(v_2-p_2)\,
\delta^{(1)}_{\mathrm{D}}(M'-M_{i})\,
\delta^{(1)}_{\mathrm{D}}(M'-M_{j})\,
\delta^{(3)}_{\mathrm{D}}(\bd{y}_1-\bd{x}_i)\,
\delta^{(1)}_{\mathrm{D}}(p_1-v_i)\,
\delta^{(3)}_{\mathrm{D}}(\bd{y}_2-\bd{x}_j)\,
\delta^{(1)}_{\mathrm{D}}(p_2-v_j)
\Big\rangle. \label{eq:fgfc_full}
\eeqa
The summation in Eq.~(\ref{eq:fgfc_full}) can be decomposed into two parts. One is the summation for the case of $i=j$, and another comes from the pairs with $i\neq j$.
The former is referred to as one-halo term, while the latter is so called two-halo term.
The one-halo term in Eq.~(\ref{eq:fgfc_full}) is then given by
\beqa
&&\Big\langle 
\int \mathrm{d}M \,
\mathrm{d}\bd{y}_1\,  \mathrm{d}\bd{y}_2\,
\mathrm{d}p_1\,  \mathrm{d}p_2\, 
\sum_{i} N(M) \, S(M)\, 
u_{\mathrm{g}}(\bd{x}_1-\bd{y}_1)\,
\delta^{(3)}_{\mathrm{D}}(\bd{x}_2-\bd{y}_2)\,
w_{\mathrm{g}}(v_1-p_1\, |\, \bd{x}_1-\bd{y}_1)\,
\delta^{(1)}_{\mathrm{D}}(v_2-p_2) \nonumber \\
&&
\quad \quad \quad \quad
\quad \quad \quad \quad
\quad \quad \quad \quad
\quad \quad \quad \quad
\times\, 
\delta^{(1)}_{\mathrm{D}}(M-M_{i})\,
\delta^{(3)}_{\mathrm{D}}(\bd{y}_2-\bd{x}_i)\,
\delta^{(1)}_{\mathrm{D}}(p_2-v_i)\,
\delta^{(3)}_{\mathrm{D}}(\bd{y}_1-\bd{y}_2)\,
\delta^{(1)}_{\mathrm{D}}(p_1-p_2)
\Big\rangle \nonumber \\
&=& \int \mathrm{d}M \, \frac{\mathrm{d}n}{\mathrm{d}M}\, N(M)\, S(M)\, P_{1}(v_2, M)\,
u_{\mathrm{g}}(\bd{x}_{1}-\bd{x}_{2}) \, w_{\mathrm{g}}(v_1-v_2\, |\, \bd{x}_{1}-\bd{x}_{2}),
\eeqa
where $\mathrm{d}n/\mathrm{d}M$ is the halo mass function
and $P_1$ represents the one-point probability distribution function of the halo velocity. In above, we use
\beqa
\Big\langle \sum_{i} \delta^{(1)}_{\mathrm{D}}(M-M_{i})\,
\delta^{(3)}_{\mathrm{D}}(\bd{y}-\bd{x}_i)\,
\delta^{(1)}_{\mathrm{D}}(p-v_i) \Big\rangle
\equiv \frac{\mathrm{d}n}{\mathrm{d}M}\, P_{1}(p, M).
\eeqa
To derive a simple form of the two-halo term, we use the fact that
\beqa
\Big\langle \sum_{i\neq j} 
\delta^{(1)}_{\mathrm{D}}(M-M_{i})\,
\delta^{(1)}_{\mathrm{D}}(M'-M_{j})\,
\delta^{(3)}_{\mathrm{D}}(\bd{y}_1-\bd{x}_i)\,
\delta^{(3)}_{\mathrm{D}}(\bd{y}_2-\bd{x}_j)\,
\delta^{(1)}_{\mathrm{D}}(p_1-v_i) \, 
\delta^{(1)}_{\mathrm{D}}(p_2-v_i)
\Big\rangle
&\equiv& 
\frac{\mathrm{d}n}{\mathrm{d}M}\, \frac{\mathrm{d}n}{\mathrm{d}M'}
\left[1+\xi_{\mathrm{hh}}(\bd{y}_{1}-\bd{y}_2, M, M')\right]
\nonumber \\
&&\times\,
P_{\mathrm{hh}}(p_1-p_2, \bd{y}_{1}-\bd{y}_2, M, M')
\nonumber \\
&&\times\,
P_{1}(p_2, M'), \label{eq:2pcf_phase_space_def}
\eeqa
where $\xi_{\mathrm{hh}}$ is the two-point correlation function of halos in real space, and $P_{\mathrm{hh}}$ represents the pairwise velocity distribution for two different halos.
Using Eq.~(\ref{eq:2pcf_phase_space_def}), we write the two-halo term as
\beqa
\int \mathrm{d}M \, \mathrm{d}M'\, \mathrm{d}\bd{y}\, \mathrm{d}p\, 
\frac{\mathrm{d}n}{\mathrm{d}M}\, \frac{\mathrm{d}n}{\mathrm{d}M'}
N(M)\, S(M')\, \left[1+\xi_{\mathrm{hh}}(\bd{y}-\bd{x}_2, M, M')\right]\, u_{\mathrm{g}}(\bd{x}_1-\bd{y})\, 
w_{\mathrm{g}}(v_1 - p\, |\, \bd{x}_1-\bd{y})
P_{\mathrm{hh}}(p-v_2, \bd{y}-\bd{x}_2, M, M')\, P_{1}(v_2, M').
\eeqa
Hence, the halo-model expression of Eq.~(\ref{eq:pairs_phase_space_def}) is given by
\beqa
{\cal F}_{\mathrm{gc}}(\bd{r}\, |\, v) &=& 
{\cal F}_{\mathrm{gc,1h}}(\bd{r}\, |\, v)+{\cal F}_{\mathrm{gc,2h}}(\bd{r}\, |\, v), \\
{\cal F}_{\mathrm{gc,1h}}(\bd{r}\, |\, v) &=& 
\int \mathrm{d}M \, \frac{\mathrm{d}n}{\mathrm{d}M}\, N(M)\, S(M)\, u_{\mathrm{g}}(\bd{r}) \, w_{\mathrm{g}}(v), \label{eq:pairs_phase_space_1h} \\ 
{\cal F}_{\mathrm{gc,2h}}(\bd{r}\, |\, v) &=& 
\int \mathrm{d}M \, \mathrm{d}M'\, \mathrm{d}\bd{y}\, \mathrm{d}p\, 
\frac{\mathrm{d}n}{\mathrm{d}M}\, \frac{\mathrm{d}n}{\mathrm{d}M'}
N(M)\, S(M')\, \left[1+\xi_{\mathrm{hh}}(\bd{y}-\bd{x}_2, M, M')\right]\, u_{\mathrm{g}}(\bd{x}_1-\bd{y})
\nonumber \\
&&
\quad \quad \quad \quad
\quad \quad \quad \quad
\quad \quad \quad \quad
\quad \quad \quad \quad
\quad \quad \quad \quad
\quad \quad \quad \quad
\times\, 
w_{\mathrm{g}}(v_1 - p\,|\,\bd{x}_1-\bd{y})
P_{\mathrm{hh}}(p-v_2, \bd{y}-\bd{x}_2, M, M').
\label{eq:pairs_phase_space_2h}
\eeqa
In this paper, 
we use the following approximation for the two-halo term:
\beqa
u_{\mathrm{g}}(\bd{x}-\bd{y}) \simeq \delta^{(3)}_{\mathrm{D}}(\bd{x}-\bd{y}).
\eeqa
Then, Eq.~(\ref{eq:pairs_phase_space_2h}) reduces to
\beqa
{\cal F}_{\mathrm{gc,2h}}(\bd{r}\, | \, v) \simeq 
\int \mathrm{d}M \, \mathrm{d}M'\,
\frac{\mathrm{d}n}{\mathrm{d}M}\, \frac{\mathrm{d}n}{\mathrm{d}M'}
N(M)\, S(M')\, \left[1+\xi_{\mathrm{hh}}(r, M, M')\right]\, 
\left[w_{\mathrm{g}}\otimes P_{\mathrm{hh}}\right](v, \bd{r}, M, M'),
\eeqa
where
\beqa
\left[w_{\mathrm{g}}\otimes P_{\mathrm{hh}}\right](v_1-v_2, \bd{r}, M, M') \equiv \int \mathrm{d}p\,
w_{\mathrm{g}}(v_1 - p\, |\, \bd{r}=\bd{0}) \, P_{\mathrm{hh}}(p-v_2, \bd{r}, M, M').
\eeqa

\subsubsection{More realistic scenarios}\label{subsubsec:offcenter_HOD}

Galaxies in the standard halo model are commonly separated into two types, centrals and satellites.
For the central galaxies, we assume that they reside in the centre of their host dark matter halos and individual host halos can have a single central galaxy at most.
For the satellite galaxies, we populate satellite galaxies to a halo only when a central galaxy exists.
In this case, the phase-space density of galaxies is written as
\beqa
f_{\mathrm{g}}(\bd{x}, v_{\parallel}) &=& 
f_{\mathrm{g, cen}}(\bd{x}, v_{\parallel}) +
f_{\mathrm{g, sat}}(\bd{x}, v_{\parallel}), \label{eq:fg_HOD} \\
f_{\mathrm{g, cen}}(\bd{x}, v_{\parallel}) &=& \sum_{i} N_{\mathrm{cen}}(M_{i}) \delta^{(3)}_{\mathrm{D}}(\bd{x}-\bd{x}_{i})
\delta^{(1)}_{\mathrm{D}}(v_{\parallel}-v_{\parallel, i}), \label{eq:f_cen} \\
f_{\mathrm{g, sat}}(\bd{x}, v_{\parallel}) &=& \sum_{i} N_{\mathrm{sat}}(M_{i}) u_{\mathrm{sat}}(\bd{x}-\bd{x}_{i}\,|\, M)
w_{\mathrm{sat}}(v_{\parallel}-v_{\parallel, i}\, |\, \bd{x}-\bd{x}_{i}, M), \label{eq:f_sat}
\eeqa
where $N_{\mathrm{cen}}(M)$ is the probability distribution function finding a central in a halo of $M$, 
$N_{\mathrm{sat}}(M)$ represents the number of satellites in the halo of $M$, $u_{\mathrm{sat}}(\bd{x}|M)$ is the number density profile of satellites, and
$w_{\mathrm{sat}}(v|\bd{x},M)$ is the distribution function of satellite velocity in the rest frame of dark matter halos.

We then arrive at
\beqa
{\cal F}_{\mathrm{gc}}(\bd{r}\, |\, v) &=&
{\cal F}_{\mathrm{cen-cl}}(\bd{r}\, |\, v) +
{\cal F}_{\mathrm{sat-cl}}(\bd{r}\, |\, v) \\
{\cal F}_{\mathrm{cen-cl}}(\bd{r}\, |\, v) &=& 
\int \mathrm{d}M \, \mathrm{d}M'\,
\frac{\mathrm{d}n}{\mathrm{d}M}\, \frac{\mathrm{d}n}{\mathrm{d}M'}
N_{\mathrm{cen}}(M)\, S(M')\, \left[1+\xi_{\mathrm{hh}}(r, M, M')\right]\, P_{\mathrm{hh}}(v, \bd{r}, M, M'), \label{eq:cen_cen} \\
{\cal F}_{\mathrm{sat-cl}}(\bd{r}\, |\, v) &=& 
\int \mathrm{d}M \, \frac{\mathrm{d}n}{\mathrm{d}M}\, N_{\mathrm{sat}}(M)\, S(M)\,
u_{\mathrm{sat}}(\bd{r}|M) \, w_{\mathrm{sat}}(v|\bd{r},M) \nonumber \\
&&+
\int \mathrm{d}M \, \mathrm{d}M'\,
\frac{\mathrm{d}n}{\mathrm{d}M}\, \frac{\mathrm{d}n}{\mathrm{d}M'}
N_{\mathrm{sat}}(M)\, S(M')\, \left[1+\xi_{\mathrm{hh}}(r, M, M')\right]\, 
\left[w_{\mathrm{sat}}\otimes P_{\mathrm{hh}}\right](v, \bd{r}, M, M'), \label{eq:sat_cen}
\eeqa
where we adopt the following notation:
\beqa
\left[w_{\mathrm{X}}\otimes P_{\mathrm{hh}}\right](v_1-v_2, \bd{r}, M, M') &\equiv& 
\int \mathrm{d}p\,
w_{\mathrm{X}}(v_1 - p\, |\, \bd{r}=\bd{0}, M) \, P_{\mathrm{hh}}(p-v_2, \bd{r}, M, M').
\eeqa

\subsection{Model specification}\label{subsec:model_spec}

Here we specify the key ingredients for our halo model of the stacked phase-space density of ACReS galaxies in and around the LoCuSS clusters. Table~\ref{tb:params} provides a short summary of our model for the LoCuSS clusters and the ACReS galaxies.

For the LoCuSS clusters, we have two ingredients, (i) mass density profiles around clusters and 
(ii) the probability distribution of cluster masses. 
We assume that the mass density in individual clusters follows a spherical symmetric NFW profile. 
We also evaluate the mass distribution of the LoCuSS clusters by using a log-normal model with the information of observed lensing masses in \citet{2016MNRAS.461.3794O}. For details, see Section~\ref{subsubsec:LoCUSS_model}.

For the ACReS galaxies, we have three ingredients, (i) the HOD, (ii) number density profiles of satellites in the LoCuSS clusters, and (iii) velocity distribution functions of satellites in the clusters. For the ACReS HOD, we basically use the observational results for photometric galaxies in the Subaru Hyper Suprime Cam survey \citep{2020ApJ...904..128I}, but calibrate the mass dependence in the satellite HOD by fitting the number density profile of the ACReS galaxies around individual LoCuSS clusters. Throughout this paper, we assume the number density profile of the ACReS galaxies in clusters can be expressed as a spherical symmetric NFW profile, while we allow the concentration parameter to be different from the counterpart in underlying mass density.
On the velocity distribution of satellites, we adopt a Gaussian velocity distribution.
We set the dispersion in the Gaussian distribution by assuming dynamical equilibrium in the LoCuSS clusters. We also include the velocity offset in the Gaussian distribution arising from gravitational redshifts of satellite galaxies in clusters. We refer Secion~\ref{subsubsec:ACReS_real_space} and \ref{subsubsec:ACReS_velocity_space} for details about our modelling of the ACReS galaxies.

\begin{table*}
\caption{
	The list of parameters involved in the model for the LoCuSS clusters and the ACReS galaxies in this paper. In the fifth column, we show the range of each parameter in our likelihood analysis (see Section~\ref{subsec:model_spec})
	\label{tb:params}
	}
\scalebox{0.80}[0.80]{
\begin{tabular}{@{}llccc}
\hline
\hline
Name & Physical meaning & Reference & Fiducial value & Range \\ 
\hline
LoCuSS clusters & & & & \\
\hline
$\Delta\log M$ & A systematic bias in the distribution of cluster lensing masses & Eq.~(\ref{eq:mass_bias}) & 0.00 & [$\log(0.96)$ : $\log(1.04)$]\\
$c_{\mathrm{m}}$ & A concentration parameter in the cluster mass profile (assuming an NFW profile) & Eq.~(\ref{eq:NFW}) & \citet{2015ApJ...799..108D} & Fixed \\
\hline
ACReS galaxies & & & & \\
\hline
$N_{\mathrm{cen}}$ & A halo occupation distribution (HOD) for centrals & Eq.~(\ref{eq:N_cen}) & \citet{2020ApJ...904..128I} & Fixed \\
$N_{\mathrm{sat}}$ & A HOD for satellites & Eq.~(\ref{eq:N_sat}) & \citet{2020ApJ...904..128I} (expect for the slope $\alpha_\mathrm{sat}$) & Fixed \\
$\alpha_{\mathrm{sat}}$ & the slope in the satellite HOD & Eq.~(\ref{eq:N_sat}) & 1.441 & [1.20 : 1.60] \\
$u_{\mathrm{sat}}$ & The number density profile of satellites & Eq.~(\ref{eq:u_sat}) 
& NFW profile & -- \\
${\cal R}$ & A concentration parameter for satellites in unit of $c_{\mathrm{m}}$ & Eq.~(\ref{eq:gal_conc}) & 0.26 & [0.20 : 0.30] \\
$w_{\mathrm{sat}}$ & The velocity distribution of satellites & Eq.~(\ref{eq:w_sat}) 
& Gaussian & -- \\
$\alpha_{\mathrm{v}}$ & The velocity bias of satellites w.r.t the solution of spherical Jeans eq. & Eq.~(\ref{eq:sigma_sat}) & 1.0 & [0.50 : 1.50]\\
$\beta$ & An anisotropic measure of satellite orbits & Eq.~(\ref{eq:Jeans}) & 0.0 & [-0.5 : 0.5]\\
$\alpha_\mathrm{mean}$ & The amplitude in gravitational redshifts & Eq.~(\ref{eq:grav_z}) & 1.0 & [0.5 : 2.0]\\
\hline
\end{tabular}
}
\end{table*}

\subsubsection{LoCuSS clusters}\label{subsubsec:LoCUSS_model}

The mass selection function $S(M)$ is the most important part for our statistical modelling of galaxy clusters.
The halo masses for individual clusters have been estimated from the weak lensing analyses in \citet{2016MNRAS.461.3794O}.
We assume that the weak lensing mass provides an estimator of the true mass $M$ and the probability of observing $M_{\mathrm{obs}}$ given the true mass $M$ takes a log-normal form of
\beqa
\mathrm{Prob}(M_{\mathrm{obs}}|M) = \frac{1}{M\, \ln 10} \frac{1}{\sqrt{2\pi}\sigma_{\log M, \mathrm{each}}}
\exp\left\{-\frac{1}{2}\left(\frac{\log M_{\mathrm{obs}}-\log M}{\sigma_{\log M, \mathrm{each}}}\right)^2\right\}, \label{eq:mass_lognormal}
\eeqa
where $\sigma_{\log M, \mathrm{each}}$ represents a typical uncertainty of the observed mass for each cluster. 
We adopt $\sigma_{\log M, \mathrm{each}} = \sqrt{(\sigma^2_{+} + \sigma^2_{-})/2}/M_{\mathrm{best}}/\ln 10$
where $M_{\mathrm{best}}$ is the best-fit lensing mass,
$\sigma_{+}$ and $\sigma_{-}$ are the upper and 
lower errors of the lensing mass estimate, respectively (see Table~\ref{tab:cluster}).
Hence, the mass distribution of the LoCuSS clusters can be computed as
\beqa
\mathrm{Prob}(M) = 
{\cal P}_0\, \sum_{i=1}^{N_\mathrm{cl}} \int \mathrm{d}M_\mathrm{obs} \, \mathrm{Prob}(M_\mathrm{obs}|M, i)\,  \delta^{(1)}_\mathrm{D}(M_\mathrm{obs}-M_\mathrm{best,i}),
\label{eq:mass_dist_model}
\eeqa
where $\mathrm{Prob}(M_\mathrm{obs}|M, i)$ is given by Eq.~(\ref{eq:mass_lognormal}) for the $i$-th cluster, 
$M_\mathrm{best,i}$ is the best-fit lensing mass for the $i$-th cluster,
$N_{\mathrm{cl}}=23$ is the number of clusters of interest,
and the normalization ${\cal P}_0$ is set by the condition of 
$\int \mathrm{d}M\, \mathrm{Prob}(M) = 1$.
Note that the mass distribution of $\mathrm{Prob}(M)$ is 
equivalent to the term of $S(M)\, \mathrm{d}n/\mathrm{d}M$ in our halo model.
\citet{2016MNRAS.461.3794O} have constrained possible 
systematic errors in the weak lensing masses of LoCuSS clusters 
to be less than 4\%.
To include this possible bias in our model, 
we shift the mass distribution as
\beqa
\mathrm{Prob}(M) \rightarrow 
\mathrm{Prob}\left(10^{\log M -\Delta \log M}\right),
\label{eq:mass_bias}
\eeqa
where $\Delta \log M$ is a free parameter in our model.

In this paper, we set the centre of each LoCuSS cluster to be the angular position of the brightest cluster galaxy (BCG). 
Although the position of the BCG may be 
different from the centre of its host halos in practice,
\citet{2010PASJ...62..811O} 
carefully examined a possible off-centring effect by studying the lensing signals with various centre proxies such as the X-ray peak, and concluded that the off-centring should be well within 100 kpc in radius. 
Therefore we ignore the off-centring effect throughout this paper.
Furthermore, we assume that the mass density distribution in the LoCuSS clusters can be expressed by a spherically-symmetric, truncated NFW profile \citep{1996ApJ...462..563N, 1997ApJ...490..493N}:
\beqa
\rho_{\mathrm{m}}(r) = \frac{\rho_{s}}{\left(r/r_s\right)\left(1+r/r_s\right)^2} \Theta(r-r_{\mathrm{200b}}), \label{eq:NFW}
\eeqa
where 
$r$ is a radius from the cluster centre,
$\Theta(x)$ is the Heaviside step function, 
$\rho_s$ and $r_s$ are a scaled density and radius, respectively.
The scaled density $\rho_s$ is given by our definition of
the spherical overdensity mass of $M_{\mathrm{200b}}$,
while the scaled radius is set by the model in \citet{2015ApJ...799..108D}.
Note that \citet{2015ApJ...799..108D} provides the prediction 
for $r_{\mathrm{200c}}/r_s$, where $r_{\mathrm{200c}}$ is the spherical overdensity radius with respect to 200 times 
the critical density in the universe.
For the conversion between $r_{\mathrm{200c}}$ and $r_{\mathrm{200b}}$, we use the fitting formula in \citet{Hu:2002we}.

\subsubsection{ACReS galaxies in real space}\label{subsubsec:ACReS_real_space}

For the ACReS galaxies, there are four quantities to specify their statistical properties as in Eqs.~(\ref{eq:fg_HOD})-(\ref{eq:f_sat}).
The number of galaxies in a halo of $M$ is determined 
by the halo occupation distribution (HOD) 
of $N_{\mathrm{cen}}(M)$ and $N_{\mathrm{sat}}(M)$.
In this paper, we adopt the observational constraints of the HOD for stellar
mass-limited galaxy samples in \citet{2020ApJ...904..128I}.
We assume that the HOD of the ACReS galaxies is given by
\beqa
N_{\mathrm{cen}}(M) &=& \frac{1}{2}\left\{ 1 + \mathrm{erf}\left(\frac{\log M -\log M_{\mathrm{cen}}}{\sigma_{\log M, \mathrm{g}}}\right)\right\}, \label{eq:N_cen}\\
N_{\mathrm{sat}}(M) &=& \left(\frac{M-M_0}{M_1}\right)^{\alpha_\mathrm{sat}} \Theta(M_0-M), \label{eq:N_sat}
\eeqa
where $M$ is the halo mass in units of $h^{-1}M_{\odot}$.
We fix the parameters to be $\log M_{\mathrm{cen}}=12.0$, $\sigma_{\log M, \mathrm{g}}=0.15$, $\log M_0 = 8.63$ and $\log M_1 = 13.5$ throughout this paper, but we allow to vary $\alpha_\mathrm{sat}$.
Note that these HOD parameters have been obtained by the clustering analysis of photometric galaxies with their stellar mass greater than $10^{10.2}\, h^{-2}M_{\odot}$ 
at $0.3\le z \le 0.55$ in the Subaru Hyper Suprime Cam (HSC) survey \citep{2020ApJ...904..128I}.
Except for $\alpha_\mathrm{sat}$, a modest difference in the stellar mass cut and 
galaxy redshifts between the ACReS and the HSC galaxy sample would not 
affect our analysis significantly, because we normalise the histogram of the cluster-galaxy pairwise velocity as varying the projected distance $r_p$.
The parameter $\alpha_\mathrm{sat}$ controls the mass dependence of the number of satellites in single clusters. This can affect the one-halo term in our model, because the stacked phase-space density is set by the histogram weighted with some function of cluster masses.

We also assume that the number density profile of satellite galaxies follows an NFW profile of
\beqa
u_{\mathrm{sat}}(r) = \frac{n_{0}}{\left(r/r_{s,\mathrm{g}}\right)\left(1+r/r_{s,\mathrm{g}}\right)^2} \Theta(r-r_{\mathrm{200b}}), \label{eq:u_sat}
\eeqa
where $r_{s, \mathrm{g}}$ is the scaled radius for the satellites
and the scaled density $n_{0}$ is set by $\int 4\pi\, r^2 \mathrm{d}r\, u_{\mathrm{sat}} = 1$.
\citet{2015ApJ...806..101H} performed a stacking analysis of the galaxy density profiles around the LoCuSS clusters.
They showed that the stacked galaxy density profile 
can be fitted by 
an NFW profile, while its best-fit scaled radius $r_{s,\mathrm{g}}$ can differ from the counterpart in the underlying cluster mass profile.
Motivated by their finding, we include the mass dependence of the scaled radius
$r_{s,{\mathrm{g}}}$ as
\beqa
r_{s,\mathrm{g}}(M) = \frac{r_{\mathrm{200b}}(M)}{{\cal R}c_{\mathrm{m}}(M)}, \label{eq:gal_conc}
\eeqa
where $c_{\mathrm{m}}(M)=r_{\mathrm{200b}}/r_s$ is the halo concentration predicted by \citet{2015ApJ...799..108D}, and ${\cal R}$ is a free parameter 
for the conversion between galaxy and mass concentration
in our model.

\begin{figure*}
\centering
\includegraphics[width=0.60\columnwidth]
{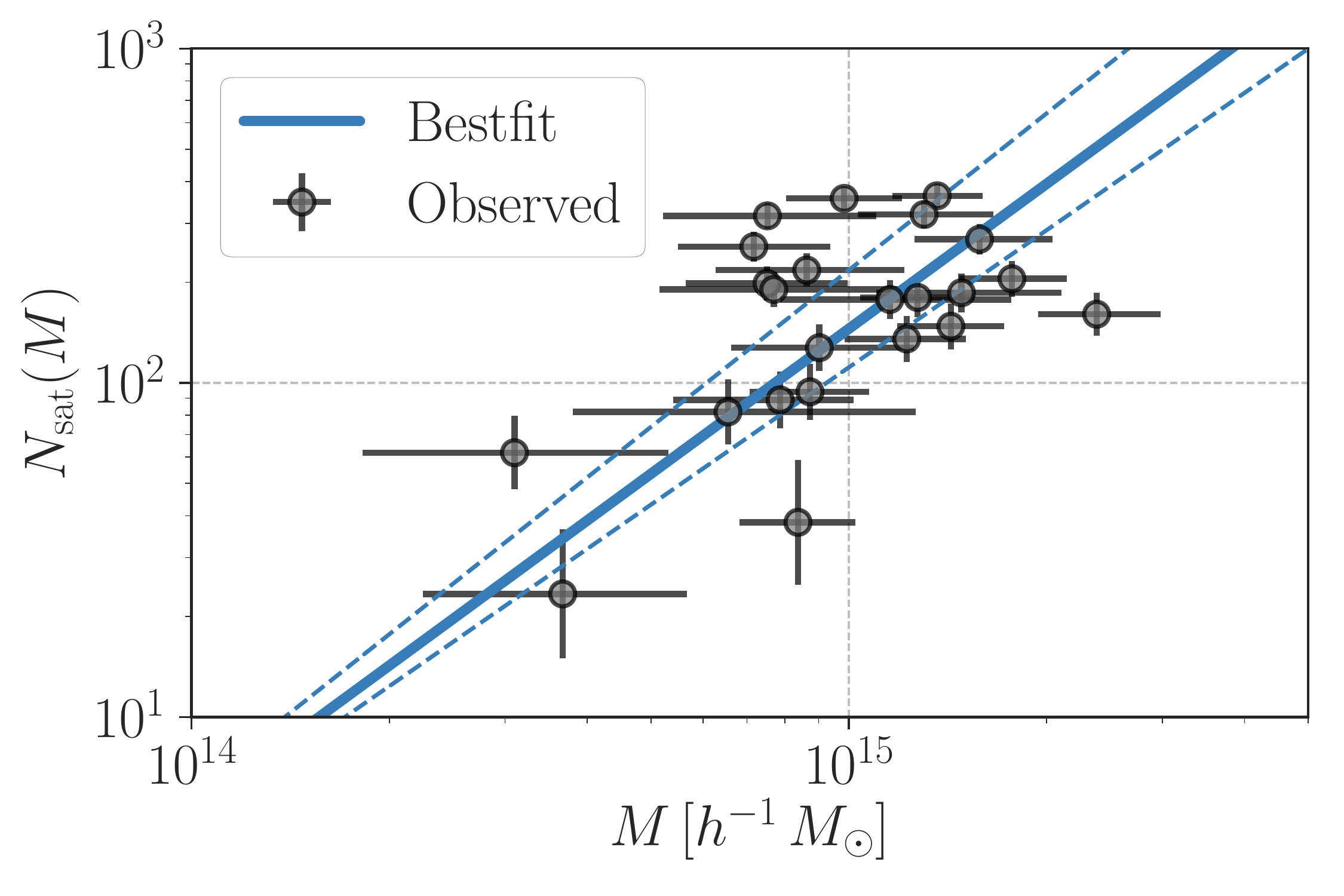}
\caption{The scatter plot of the number of ACReS satellite galaxies in individual LoCuSS clusters
as a function of cluster lensing masses. The vertical axis shows the number of satellites 
within the radius of $r_\mathrm{200b}$. We infer the number of satellites by performing a least chi-square analysis with the number density profile of the ACReS galaxies.
The blue solid line shows the best-fit model of our halo occupation distribution (Eq.~[\ref{eq:N_sat}]), while dashed lines represent a $1\sigma$-level uncertainty of the best-fit model.}
\label{fig:mass_to_Nsat}
\end{figure*}

We now study a plausible range of two parameters of $\alpha_\mathrm{sat}$ and ${\cal R}$
by using the number density profile of ACReS galaxies.
We first search for best-fit values of ${\cal R}$ and the number of satellites within the radius of $r_\mathrm{200b}$ by minimising a chi-square statistic,
\beqa
\chi^2(N_M, {\cal R})= 
\sum_{i}\frac{\left(\Sigma_{\mathrm{g, obs}}(r_{p,i}) - \Sigma_{\mathrm{g, mod}}(r_{p,i}|N_M, {\cal R})\right)^2}{\sigma^2_{\mathrm{P,i}}+\sigma^2_{\mathrm{rand,i}}},
\label{eq:chi2_indv_density}
\eeqa
where 
$N_M$ is the number of satellites within the radius of $r_\mathrm{200b}$,
$\Sigma_{\mathrm{g, obs}}(r_{p,i})$ represents the projected number density profile of the ACReS galaxies around a LoCuSS cluster at the $i$-th radius bin,
$\sigma_{\mathrm{P,i}}$ is the Poisson error at $r_{p,i}$
and $\sigma_{\mathrm{rand,i}}$ is the standard deviation derived by 10000 random samplings of the hCOSMOS galaxies.
In Eq.~(\ref{eq:chi2_indv_density}), $\Sigma_{\mathrm{g, mod}}$ presents our model prediction and is given by
\beqa
\Sigma_{\mathrm{g, mod}}(r_p\, |\, N_M, {\cal R}) = \int\, \mathrm{d}r_{\parallel}\,
N_M\, u_\mathrm{sat}\left(\sqrt{r^2_p+r^2_{\parallel}}\, |\, {\cal R}\right) + \Sigma_\mathrm{int}(r_p),
\eeqa
where 
we set the halo radius ($r_\mathrm{200b}$)
and concentration $(c_\mathrm{m})$ with the best-fit lensing mass of each LoCuSS cluster,
and $\Sigma_\mathrm{int}(r_p)$ represents the contribution from
interlopers which include uncorrelated large-scale structures with the LoCuSS clusters.
Note that we estimate the term of $\Sigma_{\mathrm{int}}$ by 10000 random samplings of the hCOSMOS galaxies as well.
In the measurement of $\Sigma_{\mathrm{g, obs}}$, 
we perform a logarithmic binning with 30 bins in the range of $0.01\le r_p \, [h^{-1}\mathrm{Mpc}]\le 10.0$, while we impose the cut of $r_p \le 3 \, h^{-1}\mathrm{Mpc}$ in Eq.~(\ref{eq:chi2_indv_density}) to mitigate possible biases due to inaccurate estimates of $\Sigma_{\mathrm{int}}$.
Figure~\ref{fig:mass_to_Nsat} shows the scatter plot in a $N_M-M$ plane by our measurements.
We infer the parameter of $\alpha_\mathrm{sat}$ from this scatter plot with a likelihood analysis.
We find $\alpha_\mathrm{sat} = 1.441^{+0.116}_{-0.077}$ with a 68\% confidence level, while details of our likelihood analysis are found in Appendix~\ref{apdx:alpha_sat}.

\begin{figure*}
\centering
\includegraphics[width=0.60\columnwidth]
{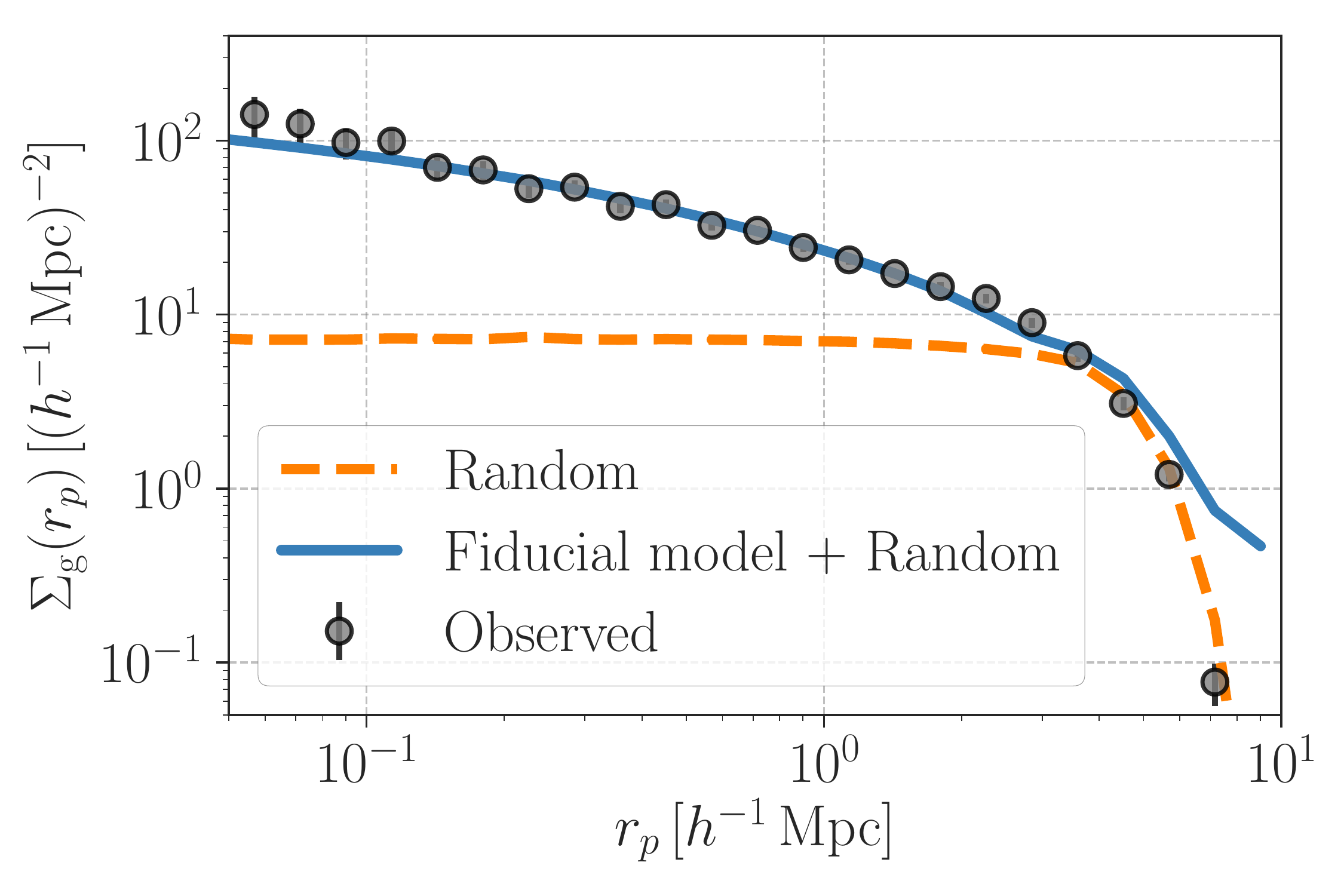}
\caption{
The stacked number density profile of the ACReS galaxies around the LoCuSS clusters. 
The grey point with error bars shows the projected number density by our measurement, while the blue solid line stands for our prediction with the best-fit value of 
${\cal R}=0.26$ (See Eq.~[\ref{eq:gal_conc}] for the definition of $\cal R$).
For comparison, we show the average projected density profile of the hCOSMOS galaxies around 10000 random points by the orange dashed line.
}
\label{fig:stacked_density_prof}
\end{figure*}

The $\chi^2$ statistic on a cluster-by-cluster basis can not place a meaningful constraint 
of ${\cal R}$.
To set a plausible range of ${\cal R}$, we perform a stacking analysis of the number density of the ACReS galaxies around 23 LoCuSS clusters. Assuming $\alpha_\mathrm{sat}=1.441$ and the cluster mass distribution of $\mathrm{Prob}(M)$ with $\Delta \log M = 0$, we can express the expected signal as
\beqa
\Sigma_{\mathrm{g, stack}}(r_p) 
&=& \bar{n}_{\mathrm{g}}
\int\, \mathrm{d}r_{\parallel}\,
\xi_{\mathrm{gc}}\left(\sqrt{r^2_p+r^2_{\parallel}}\right) + \Sigma_{\mathrm{int}}(r_p), \label{eq:stacked_density_model} \\
\xi_{\mathrm{gc}}(r) &=& 
\frac{1}{\bar{n}_{\mathrm{g}}}
\Bigg\{
\int \mathrm{d}M \, \mathrm{Prob}(M)\, N_{\mathrm{sat}}(M)\, u_{\mathrm{sat}}(r)
\nonumber \\
&&
\quad \quad \quad \quad
\quad \quad \quad \quad
+\left[\int \mathrm{d}M \, \frac{\mathrm{d}n}{\mathrm{d}M}\, \left(N_{\mathrm{cen}}(M) + N_{\mathrm{sat}}(M)\right) b_{\mathrm{L}}(M)\right]
\left[\int \mathrm{d}M \, \mathrm{Prob}(M)\, b_{\mathrm{L}}(M)\right]
\xi_{\mathrm{L}}(r) 
\Bigg\}, \label{eq:xi_gc_halomodel}\\
\bar{n}_{\mathrm{g}} &=& \int \mathrm{d}M \, \frac{\mathrm{d}n}{\mathrm{d}M}\, \left(N_{\mathrm{cen}}(M) + N_{\mathrm{sat}}(M)\right),
\eeqa
where we use the following approximation in Eq.~(\ref{eq:xi_gc_halomodel}):
\beqa
\xi_{\mathrm{hh}}(r, M, M') \simeq b_{\mathrm{L}}(M)\, b_{\mathrm{L}}(M')\, \xi_{\mathrm{L}}(r),
\eeqa
where $b_{\mathrm{L}}(M)$ is the linear halo bias,
and $\xi_{\mathrm{L}}(r)$ is the two-point correlation function of cosmic mass density by the linear perturbation theory.
To compute Eq.~(\ref{eq:xi_gc_halomodel}), we adopt the model of 
$\mathrm{d}n/\mathrm{d}M$ in \citet{2008ApJ...688..709T}
and $b_{L}$ in \citet{2010ApJ...724..878T} with the redshift being 0.2.
By performing a $\chi^2$ analysis as in Eq.~(\ref{eq:chi2_indv_density}), 
we find the best-fit value of ${\cal R}=0.26^{+0.02}_{-0.01}$ 
where we set the error bars by imposing $\chi^2({\cal R})-\chi^2(0.26)=1$.
This indicates that the ACReS satellite galaxies have a less concentrated radial profile than the underlying mass distribution on average. 
Our result is broadly consistent with the previous finding in \citet{2015ApJ...806..101H}.
Figure~\ref{fig:stacked_density_prof} shows the projected number density of the ACReS galaxies in and around the LoCuSS clusters together with our best-fit model. 
At the outermost bin, the observed stacked profile looks inconsistent 
with our model including the two-halo term.
The two halo term in $\Sigma_{\mathrm{g}}$ arises from the clustering of neighbouring halos.
This inconsistency may be explained by selection effects in the spectroscopic measurement in ACReS and/or a limited sky coverage around each LoCuSS cluster.

\subsubsection{ACReS galaxies in velocity space}\label{subsubsec:ACReS_velocity_space}

The kinematics of satellite galaxies is one of the most important parts to understand the observed phase-space density of the ACReS galaxies around the LoCuSS clusters.
In this paper, we assume that the satellites are in dynamical equilibrium within the dark matter halo, and 
that the dynamics of satellite galaxies within a given halo is
set by a spherically-symmetric system.
We then draw one-dimensional velocities of the satellites from a Gaussian
\beqa
w_{\mathrm{sat}}(v|\bd{r}) = \frac{1}{\sqrt{2\pi\, \sigma^2_{\mathrm{sat}}(r_p, r_{\parallel})}} \exp \left(-\frac{\left[v-\bar{v}(\bd{r})\right]^2}{2 \sigma^2_{\mathrm{sat}}(r_p, r_{\parallel})}\right), \label{eq:w_sat}
\eeqa
where 
$\bar{v}$ is a mean in the Gaussian,
and $\sigma_{\mathrm{sat}}(r_p, r_{\parallel})$ is the local, one-dimensional velocity dispersion along a line of sight.

For modelling $\sigma_{\mathrm{sat}}(r_p, r_{\parallel})$, we begin with the Jeans equation for a given halo of $M$
\beqa
\frac{\mathrm{d}\left(u_{\mathrm{sat}}\sigma^2_{r}\right)}{\mathrm{d}r} 
+ \frac{2\beta u_{\mathrm{sat}}\sigma^2_{r}}{r} = -u_{\mathrm{sat}}\frac{\mathrm{d}\Phi}{\mathrm{d}r}, \label{eq:Jeans}
\eeqa
where 
$u_{\mathrm{sat}}$ is the number density profile of the satellites
given by Eq.~(\ref{eq:u_sat}), $\sigma^2_{r}$ is the radial velocity dispersion, $\Phi$ is the gravitational potential in the halo of $M$, and $\beta$ is an anisotropic parameter of satellite orbits. Note that $\beta$ is defined by $1-\sigma^2_{t}/\sigma^2_{r}$ where $\sigma^2_{t}$ is the tangential velocity dispersion.
Assuming a constant $\beta$ and the gravitational potential derived by the NFW profile with the halo concentration of $c_{\mathrm{m}}$, 
we can find the solution of $\sigma^2_{r}(r)$ with the condition of $\sigma_r \rightarrow 0$ at $r\rightarrow \infty$ \citep[e.g.][]{2001MNRAS.321..155L}
\beqa
\sigma^2_{r} (r, \beta) &=& 
V^2_{\mathrm{200b}}\, g(c_{\mathrm{m}})\, s^{1-2\beta} (1+{\cal R}\, c_{\mathrm{m}}\, s)^{2}\left[I_{1}(s) - I_{2}(s)\right], \label{eq:sigma_r} \\
g(x) &=& \left[\ln(1+x)-x/(1+x)\right]^{-1}, \\
I_{1}(x) &=& \int_{x}^{\infty}\, \mathrm{d}q \frac{q^{2\beta-3}\, \ln(1+c_{\mathrm{m}}\, q)}{(1+{\cal R}\,c_{\mathrm{m}}\,q)^{2}},\\
I_{2}(x) &=& \int_{x}^{\infty}\, \mathrm{d}q
\frac{c_{\mathrm{m}}\,q^{2\beta-2}}{(1+{\cal R}\,c_{\mathrm{m}}\,q)^{2}(1+c_{\mathrm{m}}\,q)}, \\
V^2_{\mathrm{200b}} &=& \frac{G M}{r_{\mathrm{200b}}},
\eeqa
where $s = r/r_{\mathrm{200b}}$, and ${\cal R}=0.26$ is the correction factor for the concentration in the satellite density profile (see Eq.~[\ref{eq:gal_conc}]).
We then construct the model of $\sigma_{\mathrm{sat}}(r_p, r_{\parallel})$ from Eq.~(\ref{eq:sigma_r})
\beqa
\sigma_{\mathrm{sat}}(r_p, r_{\parallel})
= \alpha_{\mathrm{v}}\, \left[ 
\sigma^2_{r}(r, \beta) \left(\frac{r_{\parallel}}{r}\right)^2
+(1-\beta)\sigma^2_{r}(r, \beta) \left(\frac{r_{p}}{r}\right)^2
\right]^{1/2}, \label{eq:sigma_sat}
\eeqa
where $r^2=r^{2}_p + r^2_{\parallel}$ and $\alpha_{\mathrm{v}}$
is a free parameter in our model.
Note that we define the velocity anisotropy $\beta$ in three-dimensional space and Eq.~(\ref{eq:sigma_sat}) is valid only in the spherically-symmetric approximation.
The parameter $\alpha_{v}$ describes possible deviations from the solution of the Jean equation.
The deviation can be caused by asphericity of dark matter halos, violation of the dynamical equilibrium, and a modification of gravity in cluster regimes.
As our fiducial model, we set $\alpha_v =1$ and $\beta=0$.
When comparing the pairwise velocity histogram of the ACReS galaxies at a different $r_p$ with our model in Section~\ref{sec:results}, we will infer $\alpha_v$ and $\beta$.
Note that previous numerical simulations have shown that the velocity anisotropy of dark matter increases radially from zero in the central region to $\sim0.5$ in the outer region \citep[e.g.][]{1997ApJ...485L..13C, 1996MNRAS.281..716C, 2006NewA...11..333H}, 
while observational studies have found a marginal trend of non-zero $\beta$ on average but still consistent with $\beta=0$ within a $\sim2\sigma$ confidence level \citep[e.g.][]{2010MNRAS.408.2442W, 2013A&A...558A...1B, 2019ApJ...874...33S}.

The mean velocity in Eq.~(\ref{eq:w_sat}) can be induced by 
gravitational redshifts in galaxy clusters \citep{1995A&A...301....6C}.
A typical amplitude of the gravitational effect is expected to be $O(-10)\, \mathrm{km/s}$ \citep{2004ApJ...607..164K}, while its exact value in galaxy clusters contains 
rich information about a modification of gravity \citep[e.g.][]{2011Natur.477..567W}.
In this paper, we compute the mean velocity as
\beqa
\bar{v}(\bd{r}) = \alpha_\mathrm{mean}\frac{\Phi(0)-\Phi(\bd{r})}{c^2}, \label{eq:grav_z}
\eeqa
where $\Phi(\bd{r})$ is the gravitational potential by an NFW profile with $M$ \citep[e.g.][]{2001MNRAS.321..155L}.
Here we introduce a free parameter $\alpha_\mathrm{mean}$ to control the amplitude of the gravitational redshift effect.
As our fiducial model, we set $\alpha_\mathrm{mean}=1$, corresponding to 
the lowest-order GR prediction \citep{1995A&A...301....6C}.
It would be worth noting that higher-order effects in gravitational redshifts 
can make $\alpha_\mathrm{mean} \neq 1$ even in GR
\citep[e.g.][]{2013PhRvD..88d3013Z, 2013MNRAS.435.1278K}.
Hence, one requires careful analyses to constrain the modification of gravity with the measurement of $\alpha_\mathrm{mean}$.
In this paper, we simply regard $\alpha_\mathrm{mean}$ as a nuisance parameter.

\subsubsection{Two-halo terms}


In our halo-model framework, 
the histogram of cluster-galaxy pairwise velocities is 
given by Eq.~(\ref{eq:hist}).
This theoretical expression includes the contributions from pairwise velocity distributions for different halos with their masses of $M$ and $M'$ (denoted as 
$P_{\mathrm{hh}}(v, \bd{r}, M, M')$).
Precise modelling of $P_{\mathrm{hh}}(v, \bd{r}, M, M')$ has been developed in the literature \citep[e.g.][]{2007MNRAS.374..477T, 2013PhRvD..88b3012L, 2013MNRAS.431.3319Z, 2016MNRAS.463.3783B, 2018MNRAS.479.2256K, 2020MNRAS.498.1175C, 2021ApJ...907...38S}, while it is still difficult to predict the pairwise velocity PDF of dark matter halos over a wide range of halo masses.
In this paper, we construct numerical templates for the terms including $P_{\mathrm{hh}}(v, \bd{r}, M, M')$ based on 
the simulation data in Section~\ref{subsec:nu2gc}.
To be specific for our modelling of the pairwise velocity histogram, we require the projected phase-space density of a halo $M$ around another halo $M'$, which is expressed as
\beqa
{\cal F}_{\mathrm{hh}}(\hat{v}, r_{p}, M, M')
\equiv
\int\, \mathrm{d}r_{\parallel}\, 
\frac{\mathrm{d}n}{\mathrm{d}M}\,
\frac{\mathrm{d}n}{\mathrm{d}M'}\,
\left[1+\xi_{\mathrm{hh}}\left(\sqrt{r^{2}_p+r^{2}_{\parallel}}, M, M'\right)\right]\, 
P_{\mathrm{hh}}\left(\hat{v}-\frac{H(z)}{1+z}r_{\parallel}, r_p, r_{\parallel}, M, M' \right), \label{eq:two_halo_phase_space}
\eeqa
where $\hat{v}$ is the observed pairwise velocity,
and $H(z)/(1+z)\, r_{\parallel}$ is the Hubble flow 
due to the expansion of the universe.
The function ${\cal F}_{\mathrm{hh}}$ is computed as a function of $\hat{v}$ for a given halo pair of $M$ and $M'$ and a fixed $r_{p}$ in the simulation.
We summarise the process to estimate ${\cal F}_{\mathrm{hh}}$ with simulation data in Appendix~\ref{apdx:two_halo_sim}.
In the end, we found that the two-halo term is subdominant in the expected phase-space density at the scale less than 3 Mpc. Nevertheless, we include the contribution from the two halo terms in our analysis for the sake of completeness.

\subsection{Analysis and Information contents}

In this section, we describe the setup of our binning for the velocity histogram ${\cal H}(\hat{v}|r_p)$ and discuss the information contents in ${\cal H}(\hat{v}|r_p)$ by using our halo-model prediction.

\subsubsection{Measurements and error estimates}\label{subsubsec:measurement}

\begin{figure*}
\centering
\includegraphics[width=0.80\columnwidth]
{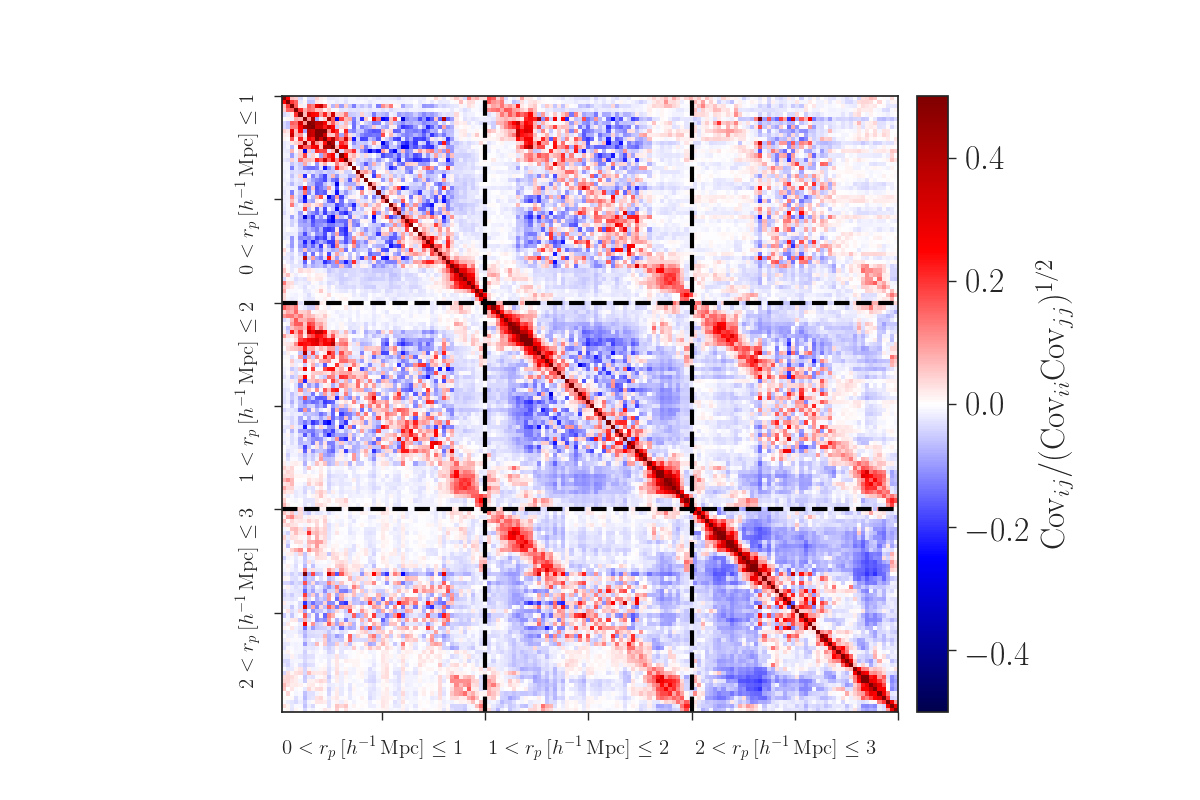}
\caption{
The cross correlation coefficient in the covariance of our phase-space density.
We estimate the covariance matrix by random sampling of the hCOSMOS galaxies as well as
simulated clusters.
}
\label{fig:cov}
\end{figure*}

To compute the histogram from the observational data, 
we perform a linear-space binning in $\hat{v}$ in the range of 
$-5000\le \hat{v}\, [\mathrm{km}\, \mathrm{s}^{-1}] \le 5000$
with the bin width of $200\, \mathrm{km}\, \mathrm{s}^{-1}$.
When constructing the velocity histogram, we also impose the selection of galaxy-cluster pairs by their projected separation length $r_{p}$. We work with a linear-space binning in $r_{p}$
and set the bin width of $\Delta r_p = 1 \, h^{-1}\mathrm{Mpc}$.
For our observational data set, we choose the outermost bin of $r_{p}$ to be $2<r_p\, [h^{-1}\mathrm{Mpc}] \le 3$.
We find that it is difficult to study the velocity histogram at $r_p > 6\, h^{-1}\mathrm{Mpc}$ with a fine bin width of $\hat{v}$,
because the sky coverage of field-of-view for individual clusters is limited and the number of available galaxies decreases at $r_p > 6\, h^{-1}\mathrm{Mpc}$ (see figure~\ref{fig:stacked_density_prof}).
In addition, interlopers can become more important at $r_p \simgt 3\, h^{-1}\mathrm{Mpc}$ as shown in Figure~\ref{fig:stacked_density_prof}.

We also estimate the statistical uncertainty in our measurement of ${\cal H}(\hat{v}|r_p)$ 
by using random sampling of the hCOSMOS galaxies and dark matter halos in the $\nu^2$GC simulation.
We first select the angular positions of 23 clusters randomly in the survey area in hCOSMOS. 
Note that we keep the redshift distribution of randomly-selected pseudo clusters same as the real counterpart.
At the position of a given pseudo cluster, we populate a cluster-sized halo from the simulation.
In this process, we randomly select the cluster-sized halo by following the mass distribution of Eq.~(\ref{eq:mass_dist_model}).
We also add (sub)halos around the cluster-sized halo.
We impose the selection cut of the (sub)halos around the simulated clusters with the \ms{halo} mass
greater than $4.7\times10^{11}\, h^{-1}M_\odot$ and separation smaller than $r_\mathrm{200b}$.
The mass cut of $4.7\times10^{11}\, h^{-1}M_\odot$ is motivated by the stellar-to-halo mass relation in \citet{2013ApJ...770...57B}.
After assigning the mock satellites from the simulation, We include the hCOSMOS galaxies around 23 mock clusters and then perform the measurement of ${\cal H}(\hat{v}|r_p)$.
We repeat this process 10,000 times and obtain 
10,000 sets of ${\cal H}(\hat{v}|r_p)$.
Finally, we evaluate the statistical error by the covariance matrix of ${\cal H}(\hat{v}|r_p)$ 
over 10,000 realisations.
Figure~\ref{fig:cov} shows the cross correlation coefficient in our covariance ${\cal H}(\hat{v}|r_p)$. 
The correlation among different bins is found to be $\simlt 0.3$ for most cases.

\subsubsection{Information contents}

\begin{figure*}
\centering
\includegraphics[width=\columnwidth]
{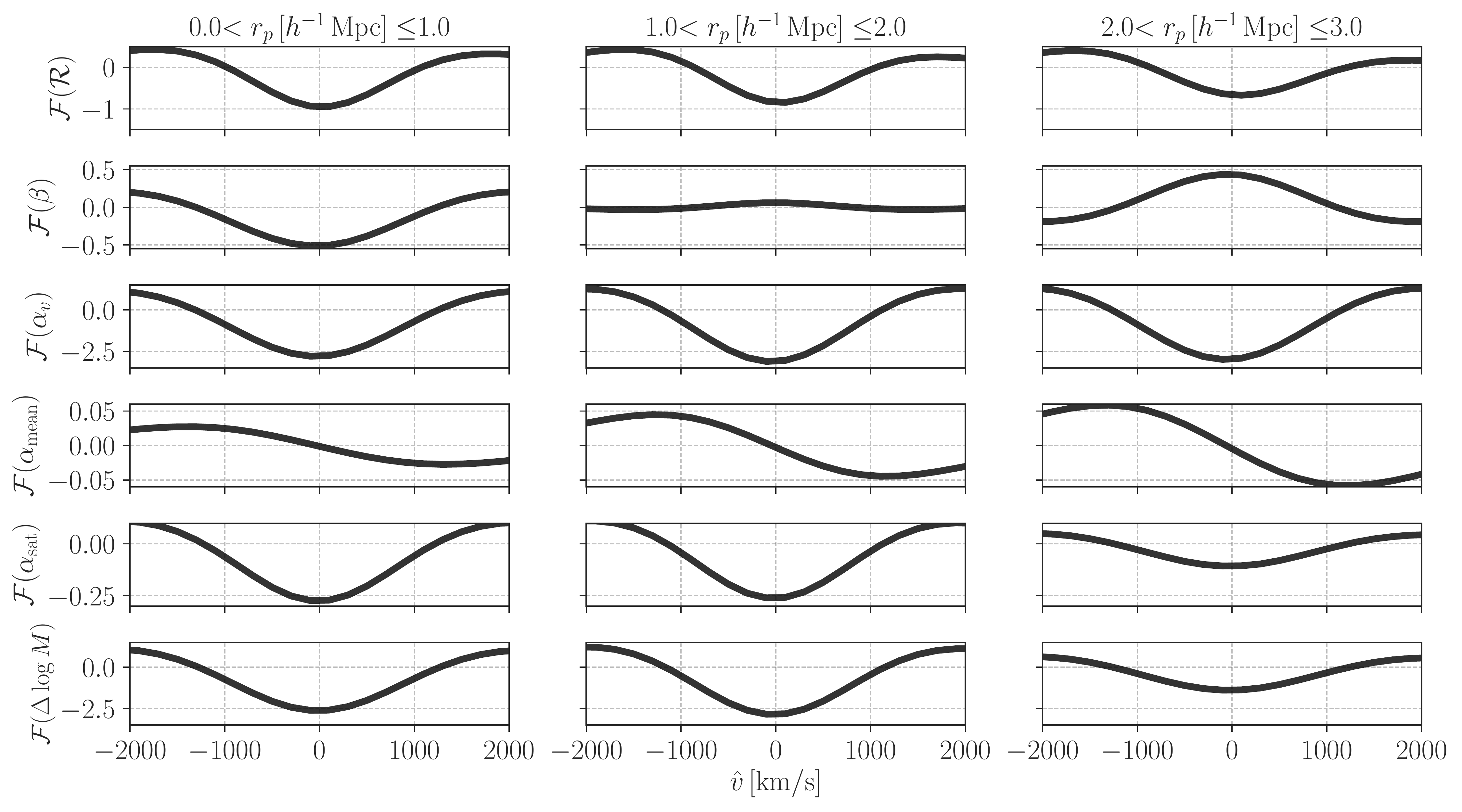}
\caption{
The dependence of the phase-space density of the ACReS galaxies 
around the LoCuSS clusters on our model parameters. 
In this figure, we define ${\cal F}(p) = 10^4 \partial {\cal H}/\partial p$, where $p$ is a parameter in our model.
From top to bottom, we show the derivative of the phase-space density with
respect to the parameter of ${\cal R}$,  $\beta$, $\alpha_v$, $\alpha_\mathrm{mean}$,
$\alpha_\mathrm{sat}$, and $\Delta \log M$, respectively.
From left to right, we show the derivative as the radius $r_p$ increases.
}
\label{fig:vPDF_param}
\end{figure*}

Before showing the main results, we summarise information contents in the phase-space density of the ACReS galaxies around the LoCuSS clusters with our halo model.
We compute the derivative of ${\cal H}(\hat{v}|r_p)$ with respect to our model parameters at
different $r_p$. The relevant parameters are listed in Table~\ref{tb:params}.
The results are shown in Figure~\ref{fig:vPDF_param}.

First of all, we emphasise that the velocity histogram 
${\cal H}(\hat{v}|r_p)$ is normalised to as 
$\int \mathrm{d}\hat{v}\, {\cal H}(\hat{v}|r_p) = 1$
for different $r_p$.
This means that our measurements of ${\cal H}(\hat{v}|r_p)$ contain less information about the number density of the ACReS galaxies.
We find that the concentration of satellite number density profile (or ${\cal R}$ in Eq.~[\ref{eq:gal_conc}]) has a small impact on our prediction of ${\cal H}(\hat{v}|r_p)$ as long as ${\cal R}$ is constrained with a level of $\sim10\%$ (see the top three panels in figure~\ref{fig:vPDF_param}).

On the other hand, ${\cal H}(\hat{v}|r_p)$ should 
have rich information about the kinematics of satellite galaxies.
The $\alpha_{v}$-dependence of ${\cal H}(\hat{v}|r_p)$ is easily understandable, 
i.e. larger $\alpha_{v}$ makes the histogram broader.
We next consider the dependence of the velocity histogram on $\beta$. 
The parameter $\beta$ controls the anisotropy of satellite orbits and it holds $\beta=0$ for the isotropic orbit, $\beta>0$ for radial orbits, and 
$\beta<0$ entails tangential orbits.
For the radial velocity dispersion $\sigma_{r}$, 
one can find larger amplitude of $\sigma_{r}$ as $\beta$ increases.
In the second top panels in figure~\ref{fig:vPDF_param}, 
we show the $\beta$ dependence becomes more complicated than that of $\sigma_r$.
For $\beta>0$, the line-of-sight velocity dispersion becomes larger than the isotropic case toward the centre of host halos, while it becomes smaller at the boundary of virial regions \citep{2001MNRAS.321..155L}.
The parameter $\alpha_\mathrm{mean}$ controls the amplitude of the gravitational redshift
and shifts ${\cal H}(\hat{v}|r_p)$ toward the direction of $\hat{v}<0$ as 
$\alpha_\mathrm{mean}$ increases.

Figure~\ref{fig:vPDF_param} also shows the 
dependence of the velocity histogram on $\Delta \log M$ and $\alpha_\mathrm{sat}$.
As $\Delta \log M$ increases, a typical halo in our cluster sample becomes more massive
and then the velocity dispersion in the stacked phase-space density becomes larger.
A similar effect can be seen when the parameter $\alpha_\mathrm{sat}$ increases,
because more massive clusters become relevant to the stacked phase-space density at larger $\alpha_\mathrm{sat}$.

\ms{It would be worth noting that an asymmetry around $v=0$ in Figure 4 can be explained by the mass dependence of the gravitational redshifts and the velocity dispersion. 
Our analytic expression of the stacked phase space density includes the integral over halo masses.
Even if the velocity PDF follows a Gaussian for single halos, 
the stacked phase space density can be (weakly) non-Gaussian after performing the mass integral. }

In summary, two parameters of $\alpha_v$ and $\beta$ can change the histogram 
at $r_p \simlt 3\, h^{-1}M_{\odot}$ predominantly.
The dependence of ${\cal H}$ on $\alpha_{v}$ and $\beta$ is different at various $r_p$.
This indicates that the phase-space analysis of galaxies around massive clusters can bring rich information about the kinematics of satellites and possible parameter degeneracies 
would be less important.
Nevertheless, we vary six different parameters to account for possible degeneracies when comparing our model with the measured ${\cal H}(\hat{v}|r_p)$.

\begin{figure*}
\centering
\includegraphics[width=\columnwidth]
{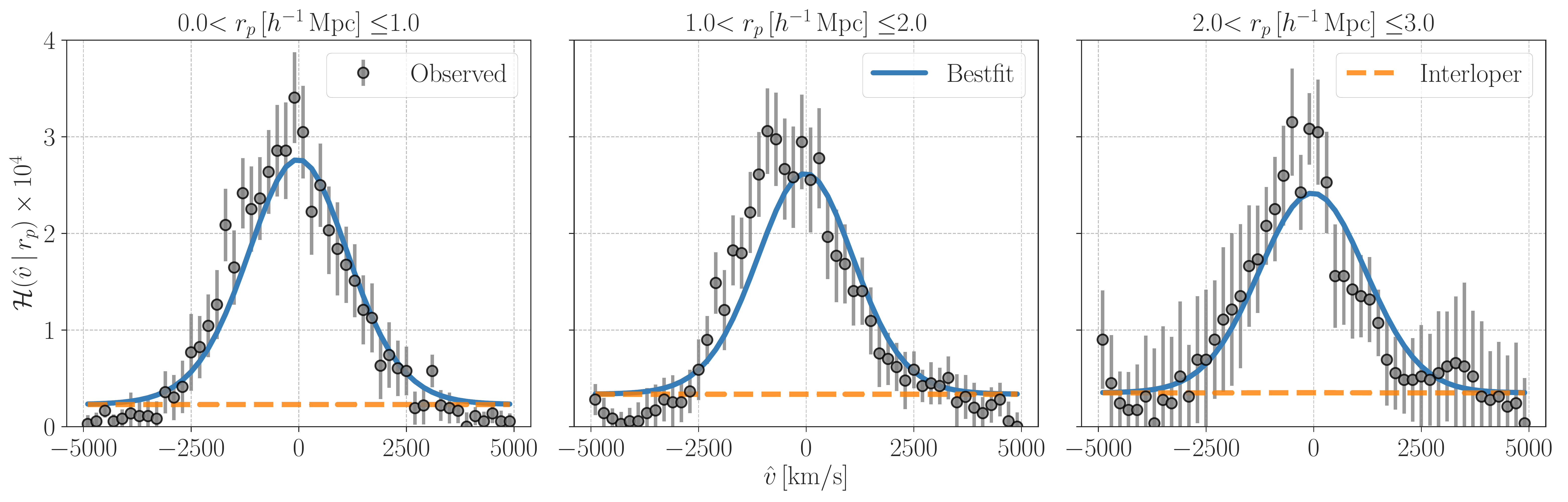}
\caption{
Comparison with the observed pairwise velocity histogram and our best-fit model. In each panel, grey points with error bars represent
the pairwise velocity histogram of ACReS galaxies around 23 LoCuSS clusters at different projected separations $r_p$.
The blue solid line shows our best-fit model,
while the orange dashed line represents the interloper term.
Our model assumes that satellite galaxies are in dynamical equilibrium in clusters
and the velocity distribution of the satellites is expressed as a Gaussian function.
The width of Gaussian velocity dispersion can be given as a spherical symmetric solution of the Jeans equation, while we introduce the mean velocity in the Gaussian distribution by considering gravitational redshifts \citep{1995A&A...301....6C}. Hence, the width of the observed velocity histogram can contain the information about the virial velocity dispersion of the ACReS galaxies in the LoCuSS clusters
and the anisotropic parameter of the galaxy orbits. See Section~\ref{subsubsec:ACReS_velocity_space} for details about our model.}
\label{fig:vPDF_data_vs_model}
\end{figure*}

\section{RESULTS}\label{sec:results}

\subsection{Inference of halo-model parameters}\label{subsec:likelihood}

We here compare the observed pairwise velocity histogram of the ACReS galaxies around the LoCuSS clusters ${\cal H}(\hat{v}|r_p)$ with our model prediction.
Our halo-based model is summarised in Section~\ref{subsec:model_spec}.
The most relevant parameters to ${\cal H}$ in our model
are the amplitude of satellite velocity ($\alpha_{v}$)
and the anisotropy parameter of satellite orbits ($\beta$),
while we account for other parameters such as 
the concentration in the number density profile of satellites ($\cal R$), 
the gravitational redshift ($\alpha_\mathrm{mean}$),
the mass dependence of the number of satellites ($\alpha_\mathrm{sat}$), 
and possible systematic mass biases in the lensing measurement ($\Delta \log M$).
To constrain those model parameters by using our measurements of ${\cal H}(\hat{v}|r_p)$,
we define a log-likelihood function as
\beqa
-2\ln {\cal L}(\bd{p}) = 
\sum_{i, j, k, l}
\left[
{\cal H}_{\mathrm{obs}}(\hat{v}_{i}\, |\, r_{p,j})
-
{\cal H}_{\mathrm{mod}}(\hat{v}_{i}\, |\, r_{p,j}, \bd{p})
\right]
\, \bd{C}^{-1}
\left[
{\cal H}_{\mathrm{obs}}(\hat{v}_{k}\, |\, r_{p,l})
-
{\cal H}_{\mathrm{mod}}(\hat{v}_{k}\, |\, r_{p,l}, \bd{p})
\right], \label{eq:log-likelihood}
\eeqa
where ${\cal H}_{\mathrm{obs}}$ is the observed velocity histogram,
${\cal H}_{\mathrm{mod}}$ is the counterpart for our model,
$\bd{p}$ represents the parameters in our model,
and $\bd{C}$ is the covariance matrix of the velocity histogram among different bins of $\hat{v}$ and $r_p$.
In the log-likelihood function, we have nine free parameters.
Six parameters among them are $\{\alpha_\mathrm{sat}, {\cal R}, \alpha_{v}, \beta, \alpha_\mathrm{mean}, \Delta \log M\}$, and
others are the interloper contribution ${\cal H}_{\mathrm{int}}$ at different $r_p$ bins (see, Eq.~[\ref{eq:hist}]).
We set a flat prior in the ranges of 
$1.20\le \alpha_\mathrm{sat} \le 1.60$,
$0.20\le {\cal R} \le 0.30$,
$0.5\le \alpha_v \le 1.5$,
$-0.5\le \beta \le 0.5$,
$0.5\le \alpha_\mathrm{mean} \le 2.0$,
and $\log(0.96)\le \Delta \log M \le \log(1.04)$
in the likelihood analysis.
To compute the log-likelihood function in our nine-parameter space, 
we use a Markov Chain Monte Carlo (MCMC) sampler of {\tt emcee} \citep{2013PASP..125..306F}.
We set 144 walkers, ran 6,250 steps to burn-in and then 56,000 steps to sample the likelihood function. We confirmed that sampling within and among chains has been converged with a level of $1\%$.

\begin{figure*}
\centering
\includegraphics[width=\columnwidth]
{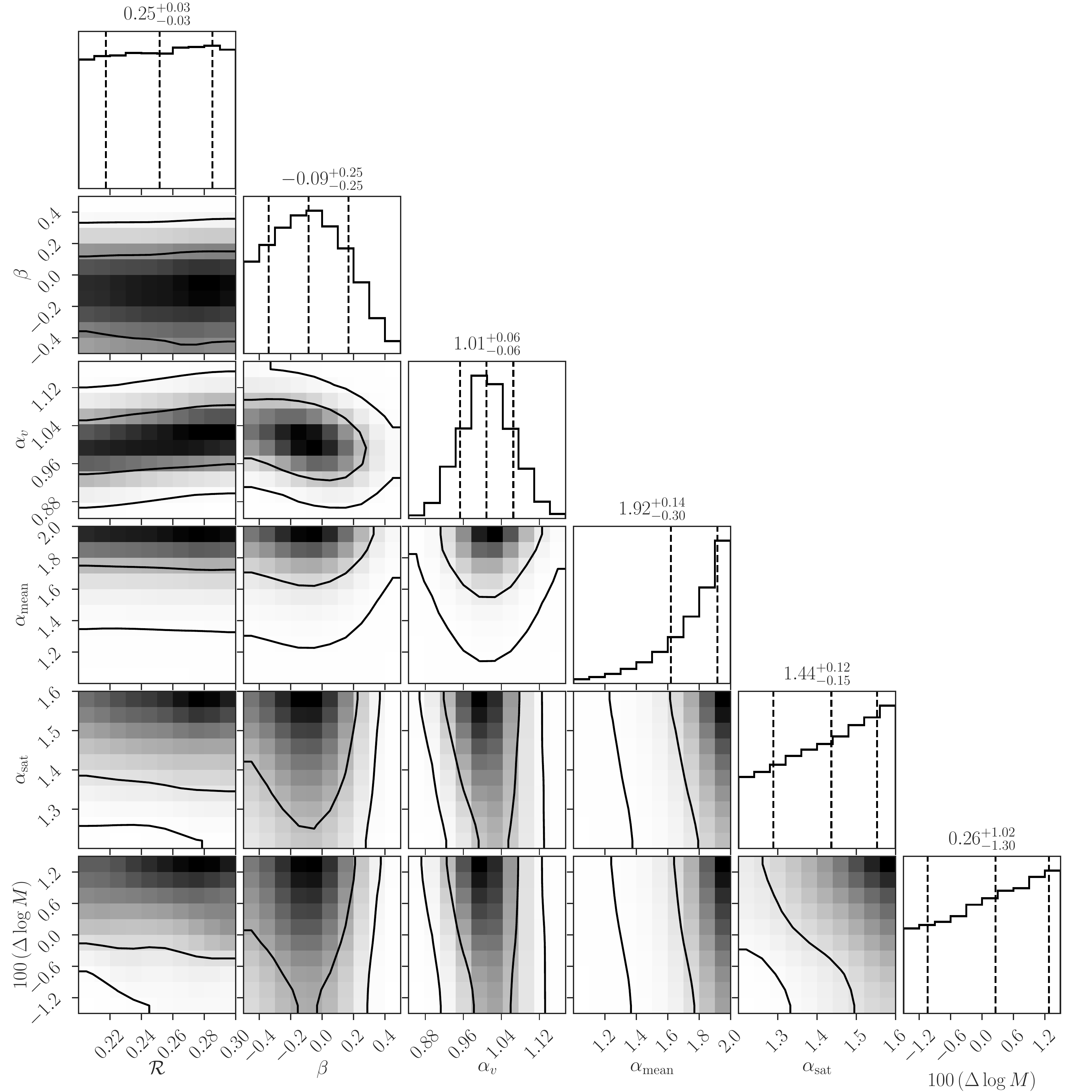}
\caption{
The posterior distribution of our halo-model parameters.
Top panels show the posterior distribution of different parameters after marginalised. In each top panel, dashed lines represent the 16th, 50th, and 86th percentiles from left to right.
Two contours in each two-dimensional plane present the 68\% and 95\% confidence levels. 
}
\label{fig:6D_params_MCMC}
\end{figure*}

Figure~\ref{fig:vPDF_data_vs_model} summarises the comparison with the observed histogram 
and the best-fit model.
Our model can provide a reasonable fit to the observed histogram in the range of $r_p \le 3\, h^{-1}\mathrm{Mpc}$.
The goodness-of-fit for our best-fit model is 
$117.8$ for the number of degrees of freedom being $150-9=141$.
The posterior distribution of our parameters is computed as 
$\mathrm{Prob}(\bd{p}) \propto {\cal L}(\bd{p}) \Pi(\bd{p})$, where 
$\Pi(\bd{p})$ represents the prior distribution.
Figure~\ref{fig:6D_params_MCMC} shows the posterior distribution of our halo-model parameters evaluated with our likelihood analysis.
After marginalising, we find $\alpha_v = 1.01^{+0.06}_{-0.06}$ and $\beta=-0.09^{+0.25}_{-0.25}$ at the 68\% confidence level.
The three parameters of ${\cal R}$, $\alpha_\mathrm{sat}$, and $\Delta \log M$
can not be constrained with our measurements.
On the other hand, we find the amplitude in the gravitational redshift to be $\alpha_\mathrm{mean} \ge 1.21$ at the 95\% confidence level.
A marginal trend of $\alpha_\mathrm{mean} \neq 1$ is mainly induced by our measurement at 
$1<r_p \, [h^{-1}\mathrm{Mpc}]\le 2$.
The histogram at $1<r_p \, [h^{-1}\mathrm{Mpc}]\le 2$ prefers a model with the mode in $\hat{v}$ being $O(-200)\, \mathrm{km/s}$.
Nevertheless, the histograms at other two $r_p$ bins 
do not present a significant non-zero velocity mode.
Because the gravitational redshift can induce more negative mean velocity at larger $r_p$,
we expect that the trend of $\alpha_\mathrm{mean} \neq 1$ may be subject to
the sample variance in our measurement\footnote{Because the standard deviation in $\hat{v}$ is typically of an order of $1000\, \mathrm{km/s}$, the Gaussian error in the average of $\hat{v}$ over 23 LoCuSS clusters is estimated to be $\sim 1000/\sqrt{23}=208\, \mathrm{km/s}$.}
or/and unknown systematic effects in the galaxy selection.


\begin{figure*}
\centering
\includegraphics[width=0.70\columnwidth]
{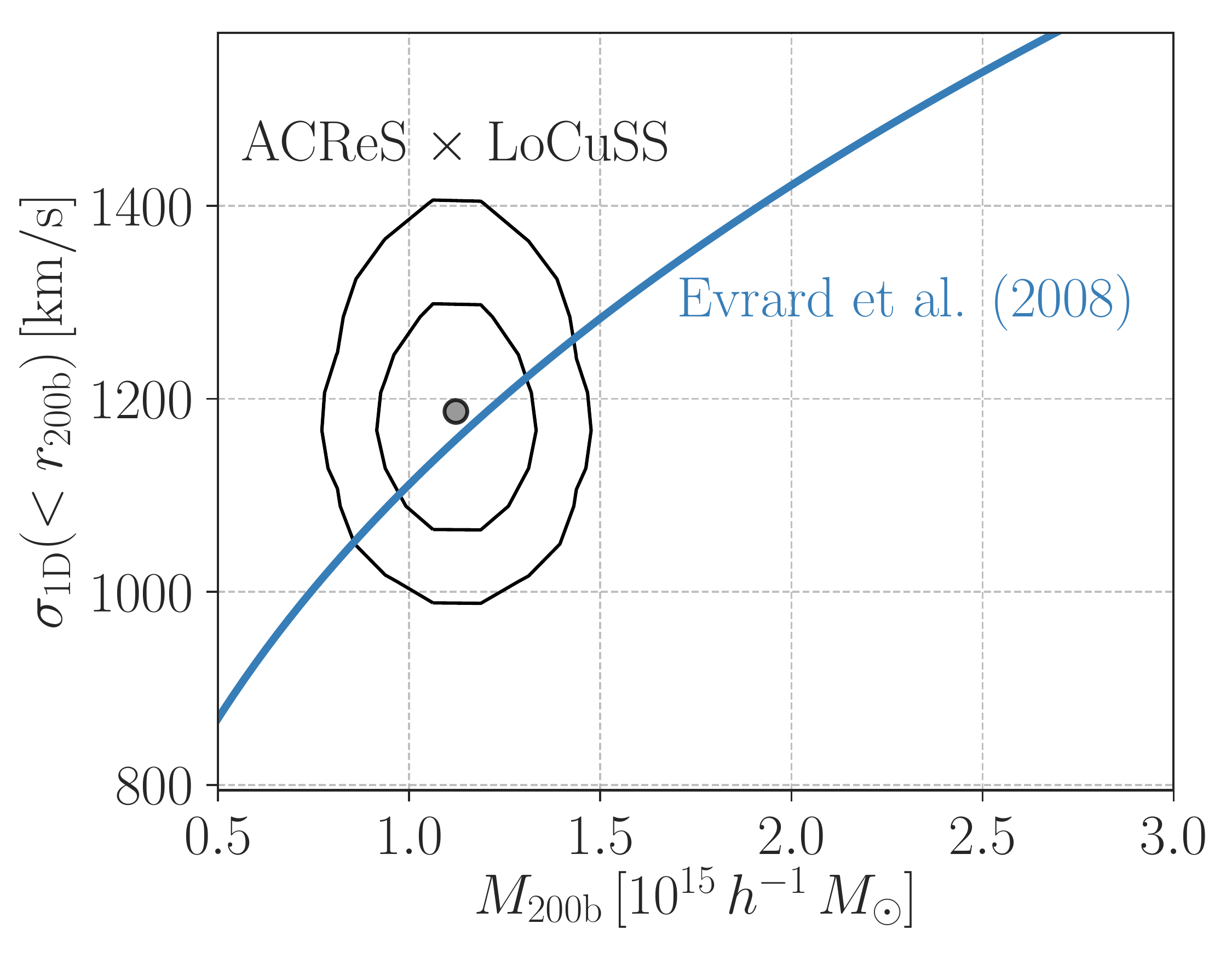}
\caption{
Internal velocity dispersion of satellite galaxies in clusters.
The grey point shows the median in the posterior distribution of $\sigma_\mathrm{1D}$ (see Eq.~[\ref{eq:sigma1D_LoCuSS}]),
while the contour is computed from the likelihood function over our model parameters.
The horizontal axis represents the average lensing mass in our LoCuSS clusters.
The two contour lines show 68\% and 95\% confidence levels.
The blue solid line shows the prediction for dark matter particles in N-body simulations under GR \citep{2008ApJ...672..122E}.
}
\label{fig:sigma1D}
\end{figure*}

Since we take a forward-modelling approach, we can predict the bulk virial scaling relation for the ACReS galaxies in the LoCuSS clusters by using our halo model.
In our halo-based model, 
the velocity dispersion within a sphere $r$ for a halo of $M$ 
is given by
\beqa
\sigma_{\mathrm{1D, J}}(<r) &=& 
\alpha_{v}\, \sqrt{\frac{3-\beta}{3}} \, \sigma_{r}(<r, \beta), \label{eq:sigma1D_Jeans} \\
\sigma^2_{r}(<r, \beta) &=& 
\frac{\int_{0}^{r} \, \mathrm{d}q\, 4\pi q^2\, u_{\mathrm{sat}}(q) \sigma^2_{r}(q, \beta)}{\int_{0}^{r} \, \mathrm{d}q\, 4\pi q^2\, u_{\mathrm{sat}}(q)},
\eeqa
where 
$u_{\mathrm{sat}}$ is set by Eq.~(\ref{eq:u_sat}),
and
$\sigma^2_{r}$ is computed as in Eq.~(\ref{eq:sigma_r}).
Taking into account the mass distribution in the LoCuSS clusters, 
we then compute 
\beqa
\sigma^2_{\mathrm{1D}}(<r_{\mathrm{200b}})
= \frac{\int \mathrm{d}\log M\, \mathrm{Prob}\left(10^{\log M-\Delta \log M}\right)\, 
N_\mathrm{sat}(M\, |\, \alpha_\mathrm{sat})\, \sigma^2_{\mathrm{1D, J}}(<r_\mathrm{200b}\, |\, M, \alpha_v, \beta, {\cal R})}{\int \mathrm{d}\log M\, \mathrm{Prob}\left(10^{\log M-\Delta \log M}\right)\, 
N_\mathrm{sat}(M\, |\, \alpha_\mathrm{sat})}, \label{eq:sigma1D_LoCuSS}
\eeqa
where $\sigma_{\mathrm{1D}}(<r_\mathrm{200b})$ depends on 
the parameters of $\alpha_\mathrm{sat}, {\cal R}, \alpha_v, \beta$ and $\Delta \log M$.
It would be worth noting that the bulk velocity dispersion 
$\sigma_{\mathrm{1D}}(<r_\mathrm{200b})$ is not a direct observable.
We are now able to infer the posterior distribution of $\sigma_{\mathrm{1D}}(<r_{\mathrm{200b}})$
with the likelihood function in Eq~(\ref{eq:log-likelihood}).
Figure~\ref{fig:sigma1D} shows the predicted velocity dispersion within a sphere of $r_{\mathrm{200b}}$ for the ACReS galaxies and LoCuSS clusters.
The horizontal axis in the figure shows the average mass over the LoCuSS clusters.
The figure highlights that the kinematics of the ACReS galaxies in the LoCuSS clusters is in good agreement with the simulation result in \citet{2008ApJ...672..122E}.
To be specific, the prediction in \citet{2008ApJ...672..122E} is given by
\beqa
\sigma_{\mathrm{DM}}(M,z) = 880\, \mathrm{km}\, \mathrm{s}^{-1}
\,\left(\frac{(1+z)^{3/2}\, M}{10^{15}\, M_{\odot}}\right)^{0.355}, \label{eq:sigma_DM}
\eeqa
where $M$ is defined as the mass of the halo enclosed in a radius containing a mean density of $200\bar{\rho}_{\mathrm{m}}$.
Our likelihood analysis sets the limit of
$\sigma_{\mathrm{1D}}(<r_{\mathrm{200b}})=1180^{+83}_{-70}\, \mathrm{km/s}$ at 
$M=1.1\times 10^{15}\, h^{-1}M_\odot$ with the 68\% confidence level,
while Eq.~(\ref{eq:sigma_DM}) gives $1145\, \mathrm{km/s}$ at $z=0.2$.
The result in figure \ref{fig:sigma1D} is also consistent with
recent findings in \citet{2018MNRAS.474.3746A}, showing that  
stellar mass-limited galaxies would exhibit almost unbiased velocity dispersions with respect to the underlying dark-matter counterparts.

\subsection{Implications to a modification of gravity}

We here discuss implications of our measurement to a modification of gravity.
In GR, the equivalence principle argues that the velocity distribution of galaxies and matter in clusters should be same as long as dark matter and galaxies are approximated as collisionless objects.
In this section, we assume that the velocity dispersion of galaxies in clusters with their mass of $M$ is simply given by Eq.~(\ref{eq:sigma_DM}) for the GR prediction.
In other words, we assume no velocity biases between galaxies and dark matter in GR. This assumption has been validated with the recent hydrodynamical simulation \citep{2018MNRAS.474.3746A}.
\ms{In addition, we assume that the weak lensing masses are not affected by the modifications of gravity.
This assumption is not valid in some of modified gravity theories \citep[e.g.][]{2015PhRvD..91f4013K}.}
\ms{When placing a limit of modified gravity theories, we vary the lensing mass in the range of $M=(1.1\pm0.1)\times10^{15}\, h^{-1}M_\odot$.
The error in the lensing mass is evaluated as the standard deviation of the average mass over 23 LoCuSS clusters.}
\ms{Hence, we take into account the $1\sigma$ uncertainty in the lensing mass estimate to constrain modified gravity theories.}

Table~\ref{tb:MG_limits} summarises limits for some of modified gravity theories by our measurements. For comparisons, we include representative limits based on different astronomical observations 
in the table.

\begin{table*}
\caption{
    The constraints of modified gravity theories given by our measurements.
    We consider three scenarios of modified Poission equation with an effective gravitational constant $G_\mathrm{eff}$, $f(R)$ gravity in \citet{2007PhRvD..76f4004H}, and the braneworld gravity proposed in \citet[][known as the DGP gravity]{2000PhLB..485..208D}. 
    In the $f(R)$ gravity, we have a free parameter $f_R$ and it represents an additional scalar field that mediates a fifth force.
    For the braneworld gravity, we have a parameter $r_c$ giving the transition scale between four- and five dimensional gravity.
    In the table, lensing represents two-point correlation analyses of cosmic shear, RSD means the redshift-space distortion in galaxy clustering analyses, $H_0$ is the prior information of the present-day Hubble parameter from local distance ladders, SNa means measurements of Hubble diagrams for supernovae, and CMB indicates measurements of temperature anisotropies in cosmic microwave backgrounds.
	\label{tb:MG_limits}
	}
\scalebox{1.00}[1.00]{
\begin{tabular}{@{}lll}
\hline
\hline
Limits & Reference & Note\\ 
\hline
Modified Poisson Equation & & \\
\hline
\ms{$0.88 \le G_\mathrm{eff}/G_\mathrm{N}\le 1.29$} & This work & $G_\mathrm{N}$ is the Newton's constant \\
$0.80 \le G_\mathrm{eff}/G_\mathrm{N}\le 1.30$ & \citet{2013MNRAS.429.2249S} & 
lensing + RSD + $H_0$ + CMB (marginalised over cosmological and nuisance params) \\
$0.74 \le G_\mathrm{eff}/G_\mathrm{N}\le 1.10$ & \citet{2019PhRvD..99h3512F} & 
lensing + RSD + CMB (marginalised over cosmological and nuisance params) \\
\hline
Hu-Sawicki $f(R)$ gravity & & \\
\hline
\ms{$\log |f_{R}| \le -4.57$} at $z=0.2$ & This work & 68\% confidence level \\
$\log |f_{R}| \le -4.22$ at $z=0$ & \citet{2014JCAP...04..013T} 
& 95\% limit, Multi-wavalength observations of the Coma cluster \\
$\log |f_{R}| \le -4.22$ at $z=0$ & \citet{2015MNRAS.452.1171W} & 95\% limit, X-ray and galaxy imaging data for 58 clusters at
$z=0.1-1.2$ \\
\hline
DGP Braneworld gravity & & \\
\hline
\ms{$r_c > 235\, h^{-1}\mathrm{Mpc}$} & This work & 68\% confidence level \\
$r_c > 6701\, h^{-1}\mathrm{Mpc}$ & \citet{2009PhRvD..80f3536L} & 95\% limit, CMB + SNa (marginalised over cosmological params)\\
$r_c > 464-870\, h^{-1}\mathrm{Mpc}$ & \citet{2013MNRAS.436...89R} & 68\% limit, RSD (fixed cosmological params, but maginalised over galaxy bias params)\\
\hline
\end{tabular}
}
\end{table*}

\subsubsection{Modified Poission equation}

In modified gravity theories, the Poisson equation is commonly modified as
\beqa
\nabla^2 \Phi = 4\pi G_\mathrm{eff}\, \bar{\rho}_\mathrm{m} \, \delta_\mathrm{m}\, a^2,
\label{eq:modified_Poission}
\eeqa
where 
$\Phi$ is the gravitational potential, 
$G_\mathrm{eff}$ is a modified gravitational constant, 
$\bar{\rho}_\mathrm{m}$ is the mean matter density,
$\delta_\mathrm{m}$ is the constrast of matter density,
and $a$ is the scale factor in the Friedmann--Lemaitre--Robertson--Walker metric.
It would be worth noting that $G_\mathrm{eff}$ can depend on time and spatial coordinates in general.
In the limit of GR, it holds that $G_\mathrm{eff} \rightarrow G_\mathrm{N}$,
where $G_\mathrm{N}$ is the Newton's constant.

The virial theorem for a collisionless system predicts that 
the galaxy velocity dispersion under Eq.~(\ref{eq:modified_Poission}) is given by \citep[e.g.][]{2010PhRvD..81j3002S}
\beqa
\frac{\sigma^2_\mathrm{1D, MG}}{\sigma^2_\mathrm{1D, GR}} = \frac{G_\mathrm{eff}}{G_\mathrm{N}},
\eeqa
where $\sigma_\mathrm{1D, MG}$ is the galaxy velocity dispersion under the modified Poisson equation, and $\sigma_\mathrm{1D, GR}$ is the GR counterpart.
Because the weak lensing measurement of the LoCuSS clusters 
can give an estimate of $\sigma_\mathrm{1D, GR}$ with Eq.~(\ref{eq:sigma_DM}), 
our measurement of $\sigma_\mathrm{1D}$ places a limit of 
\ms{$0.88 < G_\mathrm{eff}/G_\mathrm{N} < 1.29$ at the halo mass of $M=(1.1\pm0.1)\times10^{15}\, h^{-1} M_\odot$} 
with the 68\% confidence level.
Our constraint is found to be comparable to one from large-scale structures \citep[e.g.][]{2013MNRAS.429.2249S, 2019PhRvD..99h3512F, 2020JCAP...12..018G}, 
while relevant physical scales to our measurement is of an order of 
$\mathrm{Mpc}$ and much shorter than typical scales of large scale structures (10-100 $\mathrm{Mpc}$).
\ms{We here caution that our analysis still ignores possible impacts on lensing masses by the GR modification
when constraining $G_\mathrm{eff}/G_\mathrm{N}$.}

\subsubsection{$f(R)$ gravity}

$f(R)$ gravity is a class of modified gravity theories where the Lagrangian in this theory
involves an arbitrary function of the Ricci scalar $R$.
In this gravity theory, an additional scalar degree of freedom is given by
$f_R\equiv \mathrm{d}f/\mathrm{d}R$ and introduces a fifth force through its field equation.
As a representative example, we adopt the functional form of $f(R)$ as proposed in \citet{2007PhRvD..76f4004H},
\beqa
f(R) = -2\Lambda - f_{R0}\frac{\bar{R}^2_0}{R},
\eeqa
where 
$\bar{R}_0$ is the present-day Ricci scalar for the background space-time,
and $f_{R0}$ and $\Lambda$ are free parameters in this model.
The first term can be regarded as an effective cosmological constant
yielding accelerated expansion of the universe, whereas the second term controls 
the deviation from GR.
Note that the model requires $f_{R} < 0$ to be stable under perturbations \citep{2007PhRvD..76f4004H}.
The dynamical mass in this model has been investigated with a set of N-body simulations.
\citet{2018MNRAS.477.1133M} found the relation between the dynamical and lensing masses in this model can be well expressed as
\beqa
\frac{M_{\mathrm{dyn}}}{M_\mathrm{lens}} &=& \frac{7}{6} - \frac{1}{6}\tanh \left[p_1 \log \left(\frac{M_\mathrm{lens}}{M_\odot}\right)+p_2\right], \label{eq:M_dyn_fR}\\
p_1 &=& 1.503\, \log \left(\frac{|f_R(z)|}{1+z}\right) +21.64, \\
p_2 &=& 2.21,
\eeqa
where 
$M_\mathrm{dyn}$ represents the dynamical mass, and
$M_\mathrm{lens}$ is the lensing mass defined 
by a spherical overdensity mass with the enclosed mass density being 
500 times the critical density (referred to as $M_\mathrm{500c}$).
It would be worth noting that the lensing equation in this $f(R)$ model is equivalent to the GR counterpart as long as $|f_{R0}| \ll 1$ \citep[e.g.][]{2014MNRAS.440..833A}.
Because the virial theorem predicts that the potential energy in an NFW halo is proportional to $M^{5/3}_\mathrm{dyn}$ as well as $\sigma^{2}_\mathrm{1D}$ \citep[e.g.][]{2010PhRvD..81j3002S}, 
we can rewrite Eq~(\ref{eq:M_dyn_fR}) as
\beqa
\frac{\sigma^2_\mathrm{1D, fR}}{\sigma^2_\mathrm{1D,GR}}
= \Bigg\{\frac{7}{6} - \frac{1}{6}\tanh \left[p_1 \log \left(\frac{M_\mathrm{lens}}{M_\odot}\right)+p_2\right]\Bigg\}^{5/3}, \label{eq:sigma_1D_fR}
\eeqa
where $\sigma_{\mathrm{1D,fR}}$ is the galaxy velocity distribution in the $f(R)$ gravity, 
and $\sigma_\mathrm{1D,GR}$ is given by Eq.~(\ref{eq:sigma_DM}).
Using Eq.~(\ref{eq:sigma_1D_fR}), we find that our measurement of $\sigma_\mathrm{1D}$ provides an upper limit of 
\ms{$\log |f_{R}(z=0.20)| < -4.57$} at the 68\% confidence level.
When deriving this limit, we convert the lensing mass $M_\mathrm{200b}$ to $M_\mathrm{500c}$ with the halo concentration in \citet{2015ApJ...799..108D}.
Our constraint of $f_{R}$ is comparable to previous limits obtained by the Coma cluster \citep{2014JCAP...04..013T} and 58 clusters in the XMM Cluster Survey \citep{2015MNRAS.452.1171W}.
We here emphasise that our analysis measures the dynamical mass with the kinematic information of galaxies in clusters, while the previous analyses have to assume the hydrostatic equilibrium to infer $M_\mathrm{dyn}$ from measurements of gas in clusters.

\subsubsection{DGP braneworld gravity}

Dvali, Gabadadze, and Porrati (DGP) proposed a model such that our universe is a $(3+1)$-brane embedded in five-dimensional Minkowski space \citep{2000PhLB..485..208D}.
Gravity in this model becomes five-dimensional at scales larger than the crossover scale $r_c$, while it is four-dimensional at scales shorter than $r_c$.
This model has two branches of homogeneous cosmological solutions on the brane.
One is the self-accelerating branch allowing the late-time acceleration of the universe without introducing a cosmological constant.
Another is the normal branch which needs dark energy to realise the cosmic acceleration at $z\simlt1$.
In this paper, we consider the normal branch because cosmological observables in the self-accelerating branch are in conflict with the data \citep[e.g][]{2006PhLB..642..432F, 2006PhRvD..74b3004M, 2008PhRvD..78j3509F}.
The dynamical mass in the normal branch of DGP (nDGP) has been studied in \citet{2010PhRvD..81j3002S},
\ms{while the lensing mass is not affected by the modification of gravity in the DGP model \citep{2009PhRvD..80d3001S}.}
On sub-horizon scales, gravitational force at scales shorter than $r_c$ is governed by two potentials of $\Phi_\mathrm{N}$ and $\phi$, where $\Phi_\mathrm{N}$ is the gravitational potential in GR and $\phi$
represents an additional scalar potential in this model.
For a spherical symmetric halo, an effective gravitational constant in the nDGP model 
can be expressed as \citep{2010PhRvD..81f3005S, 2010PhRvD..81j3002S}
\beqa
\frac{G_\mathrm{eff, nDGP}(r,z)}{G_\mathrm{N}} &=& 
1+\frac{2}{3\beta_\mathrm{DGP}(z)}
\, g\left(\frac{r}{r_*(r,z)}\right), \label{eq:G_eff_DGP} \\
g(y) &=& y^3 \left(\sqrt{1+y^{-3}}-1\right)
\eeqa
where the parameter $\beta_\mathrm{DGP}$ is given by
\beqa
\beta_\mathrm{DGP}(z) = 1+2\, H(z)r_c\, \left(1+\frac{\dot{H}(z)}{3H^2(z)}\right),
\eeqa
where $H(z)$ is the Hubble parameter at redshift $z$ and $\dot{H}$ is the time derivative of $H(z)$.
The radius of $r_*$ in Eq.~(\ref{eq:G_eff_DGP}) is the Vainshtein radius defined as
\beqa
r_*(r,z) = \left(\frac{16G\, M(<r)r_c^2}{9\beta^2_\mathrm{DGP}(z)}\right)^{1/3},
\eeqa
where 
$M(<r)$ represents the enclosed mass of the halo within $r$.
The Vainstein radius depends on the radial coordinate $r$
and the fifth-force vanishes at $r \ll r_*$.
Hence, the observed galaxy velocity dispersion in the nDGP can be modified as
\beqa
\frac{\sigma^2_\mathrm{1D, nDGP}}{\sigma^2_\mathrm{1D, GR}}
= \frac{\int_0^{r_\mathrm{200b}} \, 4\pi\, q^2 \mathrm{d}q\, u_\mathrm{sat}(q)\, G_\mathrm{eff, nDGP}(q,z)/G_\mathrm{N}}{\int_0^{r_\mathrm{200b}}  \, 4\pi\, q^2 \mathrm{d}q\, u_\mathrm{sat}(q)}. \label{eq:sigma_1D_nDGP}
\eeqa
Using Eq.~(\ref{eq:sigma_1D_nDGP}) and our measurement of $\sigma_\mathrm{1D}$, 
we find a lower bound of \ms{$r_c > 253\, h^{-1}\mathrm{Mpc}$} with the 68\% confidence level at the halo mass of \ms{$M=(1.1\pm0.1)\times 10^{15}\, h^{-1}M_{\odot}$} and $z=0.20$.
Our constraint of $r_c$ is weaker than previous limits derived by the cosmic microwave background
and Hubble diagram of supernovae \citep[e.g.][]{2009PhRvD..80f3536L}, but comparable to one from large scale structures \citep[e.g.][]{2013MNRAS.436...89R}.
Note that the current tightest constraint of $r_c$ is mainly set by the geometric information of the universe, while our constraint is based on the gravitational interaction in galaxy clusters at relevant scales of $\sim3\, \mathrm{Mpc}$.

\section{CONCLUSIONS AND DISCUSSIONS}\label{sec:conclusion}

In this paper, we measured average histograms of pairwise velocity of galaxies in and around clusters at different projected cluster-galaxy separations, referred to as the stacked phase-space density.
We employed a halo-model forward-modelling approach for the stacked phase-space density.
In our model, we take into account realistic effects on the phase-space density 
including a wide distribution of cluster masses,
satellite galaxies in single clusters, and large-scale streaming motions 
between two different dark matter halos.
We specified key ingredients in our model by using 
the spectroscopic data of galaxies around the LoCuSS clusters \citep{2015ApJ...806..101H} 
with precise weak-lensing mass estimates \citep{2016MNRAS.461.3794O}.
We then performed the stacking analysis of phase-space density around the LoCuSS clusters and made a comparison with our model.
A likelihood analysis enables us to infer kinematics of our galaxy sample in the LoCuSS clusters.
We found that the velocity dispersion of our selected galaxies within the virial region is
constrained to be $1180^{+83}_{-70}\, \mathrm{km/s}$ at the 68\% confidence level
at the cluster mass of \ms{$(1.1\pm0.1)\times10^{15}\, h^{-1}M_\odot$}.
Our constraint of the galaxy velocity dispersion is consistent with the prediction by dark-matter-only N-body simulations under General Relativity \citep{2008ApJ...672..122E}.
Because our galaxy selection is designed to be approximately stellar mass-limited,
our finding confirms recent numerical results in \citet{2018MNRAS.474.3746A}, showing
that stellar mass-limited galaxies can be used as 
a good tracer of the gravitational potential inside clusters.

Using our measurement of the galaxy velocity dispersion in clusters with known lensing masses, 
we put a constraint of possible modifications of gravity.
If the Poisson equation at scales of Mpc can be modified with an effective gravitational constant $G_\mathrm{eff}$, our measurement can place a limit of \ms{$0.88<G_\mathrm{eff}/G_\mathrm{N}<1.29$} with the 68\% confidence level at the halo mass of \ms{$(1.1\pm0.1)\times10^{15}\, h^{-1}M_\odot$} and redshift of 0.2,
where $G_\mathrm{N}$ is the Newton's constant.
This constraint is found to be comparable to the previous limits by large-scale structures \citep[e.g.][]{2013MNRAS.429.2249S, 2019PhRvD..99h3512F, 2020JCAP...12..018G}.
Furthermore, we found that our constraint of the galaxy velocity dispersion in clusters gives an upper limit of an additional scalar field in $f(R)$ gravity to be \ms{$\log |f_{R}(z=0.2)| < -4.57$} (68\%),
where $f_{R}$ is the scalar field in the $f(R)$ gravity.
For the braneworld gravity proposed in \citet{2000PhLB..485..208D} (known as the normal branch of DGP), our measurement requires the crossover distance on the brane to be larger than \ms{$253\, h^{-1}\mathrm{Mpc}$} 
at the 68\% confidence level.
Our limits of $f(R)$ and DGP gravity models are broadly consistent with the previous results 
by measurements of intracluster medium \citep[e.g.][]{2014JCAP...04..013T, 2015MNRAS.452.1171W} as well as the large-scale structures \citep[e.g.][]{2013MNRAS.436...89R}.

Our stacking analysis in phase space uses 16071 galaxies around 23 LuCuSS clusters, allowing us to measure the bulk velocity dispersion with a 7\% level precision.
To further tighten the constraint of modified gravity, future observations of 
stacked phase-space density need to increase number of clusters as well as spectroscopic observations of galaxies around each cluster.
As the statistical uncertainty becomes smaller, we require more detailed analysis to test gravity.
Possible velocity offsets between BCGs and their host halos induced by mergers \citep[e.g.][]{2014ApJ...786...79M} and scale dependences of the velocity anisotropy $\beta$ are not included in our framework.
It would be worth noting that small spatial offsets between BCGs and their host halos do not necessarily imply small velocity offsets \citep{2011MNRAS.410..417S}.
The distribution of galaxies at the edge of cluster boundaries can be more complicated than our model predictions.
The sharp truncation in galaxy density profiles at cluster outskirts has been observed \citep[e.g.][]{2016ApJ...825...39M, 2018ApJ...864...83C, 2020PASJ...72...64M, 2020arXiv201005920B},
while infall galaxies onto clusters can have non-Gaussian velocity distributions \citep{2019MNRAS.488.4117H, 2021MNRAS.502.1041A}.
Our model can not account for these effects correctly, because the mass accretion history of individual clusters plays an important role in determining properties of galaxies at cluster outskirts \citep[e.g.][]{2015ApJ...810...36M}.
More spectroscopic observations play a central role in verifying the presence of infall galaxies 
in and around galaxy clusters.
A joint analysis with stacked weak lensing is promising to 
explore a broad range of modified gravity models \citep[e.g.][]{2015MNRAS.454.4085B, 2015JCAP...10..064T}, but a more precise estimation of statistical errors will be demanded.


\section*{acknowledgements} 

We thank the anonymous referee for providing useful comments.
This work is in part supported by MEXT KAKENHI Grant Number (18H04358, 19K14767, 20H05861). Numerical computations were in part carried out on Cray XC30 and XC50 at Center for Computational Astrophysics, National Astronomical Observatory of Japan.

\section*{Data Availability}
We have used the spectroscopic data obtained by ACReS, which will be shared on reasonable request to the authors. Other data includes the $\nu^2$GC halo catalogue, weak-lensing mass estimates of LoCuSS clusters, and the hCOSMOS catalogue. These are freely publicly available.

\appendix

\section{A likelihood analysis for the Halo Occupation distribution of satellite galaxies}\label{apdx:alpha_sat}
We here summarise how to infer the halo occupation distribution of satellites in single clusters with a likelihood analysis.
For this purpose, we use the scatter plot between the observed number of satellites and lensing masses as in Section~\ref{subsubsec:ACReS_real_space} (see Figure~\ref{fig:mass_to_Nsat}).
We assume that the average number of satellites in clusters with $M$ can be given by Eq.~(\ref{eq:N_sat}).
Fixing the parameters of $\log M_0 \, [h^{-1}M_\odot] = 8.63$ and $\log M_1 \, [h^{-1}M_\odot] = 13.5$, we perform a likelihood analysis to constrain $\alpha_\mathrm{sat}$
for the distribution in $N_M$, where $N_M$ is the observed number of satellites in a cluster.
Assuming that $N_M$ follows a Poisson distribution, we compute the likelihood function as
\beqa
{\cal L}(\alpha_\mathrm{sat}) \equiv \prod_{i=1}^{N_{\mathrm{bin}}}\frac{\lambda_{i}^{N_{\mathrm{cl},i}}\exp(-\lambda_{i})}{N_{\mathrm{cl},i}!}, \label{eq:L_poisson}
\eeqa
where 
we perform a binning in $N_M$ with the number of bins being $N_\mathrm{bin}$,
$N_{\mathrm{cl}, i}$ is the number of the LoCuSS clusters at the $i$-th bin of $N_M$, 
and $\lambda_{i}$ is an expected number of the clusters at the $i$-th bin.
We set the model of the number count of the LoCuSS clusters as a function of $N_M$ below:
\beqa
\lambda(N_M, \alpha_\mathrm{sat}) &=& \sum_{i=1}^{N_\mathrm{cl}} \int \mathrm{d}M\,
\mathrm{d}M_\mathrm{obs}\, 
\mathrm{Prob}(N_M\, |\, \alpha_\mathrm{sat}, M, i)\, \mathrm{Prob}(M_{\mathrm{obs}}\, |\, M, i)\, 
\delta^{(1)}_\mathrm{D}(M_\mathrm{obs}-M_{\mathrm{best},i}), \\
\mathrm{Prob}(N_M\, |\, \alpha_\mathrm{sat}, M, i)
&=&
\frac{1}{N\, \ln 10}\, \frac{1}{\sqrt{2\pi} \sigma_{\log N, i}}
\exp\left\{-\frac{1}{2}\left(\frac{\log N_M-\log N_\mathrm{sat}(M, \alpha_\mathrm{sat})}{\sigma_{\log N, i }}\right)^2\right\},
\eeqa
where $\mathrm{Prob}(M_\mathrm{obs}|M, i)$ is given by Eq.~(\ref{eq:mass_lognormal}) for the $i$-th cluster, 
$M_{\mathrm{best},i}$ is the best-fit weak lensing mass of the $i$-th cluster,
$N_\mathrm{sat}(M)$ is given by Eq.~(\ref{eq:N_sat}),
and $\sigma_{\log N, i}$ represents the statistical uncertainty in the measurement of $\log N_M$ for the $i$-th cluster.
Note that $\sigma_{\log N, i}$ can be obtained through the $\chi^2$ analysis on a cluster-by-cluster basis.
In Eq.~(\ref{eq:L_poisson}), we employ 
a logarithmic binning with $N_{\mathrm{bin}}=5$ and 
the bin width being $\Delta \log N_M = 1.7$.
We also set $\log N_{M,i} = 1.0 + (i-0.5) \Delta \log N_M$ for $i=1,\cdots, 5$.
We then find the best-fit parameter of $\alpha_\mathrm{sat}$ by maximising the likelihood function (Eq.~[\ref{eq:L_poisson}]).
To do so, we adopt a flat prior in 
a range of $1.20 \le \alpha_\mathrm{sat}\le 1.80$.

\section{Two halo terms for phase-space density based on numerical simulations}\label{apdx:two_halo_sim}

In this appendix, we describe how to estimate the two-halo term in our halo model of phase-space density in Eq.~(\ref{eq:two_halo_phase_space}).
Suppose that we have a catalogue of dark matter halos in a cubic box with the length on a side being $L_{\mathrm{box}}$.
To estimate Eq.~(\ref{eq:two_halo_phase_space}) from the halo catalogue, we employ a distant-observer approximation and set an axis in the simulation box to be the line-of-sight direction.
We denote $x_3$ as the line-of-sight direction.
To compute the function ${\cal F}_{\mathrm{hh}}$, we perform a binning in $\hat{v}$, $r_{p}$, $M$, and $M'$. The bin widths for four quantities are referred to as 
$\Delta \hat{v}$, $\Delta r_{p}$, $\Delta M$, 
and $\Delta M'$, respectively.
We first find a set of halos with their mass of 
$M_i-\Delta M/2 \le M < M_{i}+\Delta M/2$ 
where $M_i$ is the $i$-th mass bin.
We refer these halos as $i$-th primary sample.
We then count the number of halos with their mass of 
$M'_{j}-\Delta M'/2 \le M' < M'_{j}+\Delta M'/2$
around the $i$-th primary sample.
We perform the number count over the range of $-L_{\mathrm{box}}/2<r_{\parallel}<L_{\mathrm{box}}/2$
by imposing the cut of 
$\hat{v} = [\hat{v}_{\alpha}-\Delta \hat{v}/2, \hat{v}_{\alpha}+\Delta \hat{v}/2]$
and 
$r_{p} = [r_{p,\beta}-\Delta r_p/2, r_{p,\beta}+\Delta r_p/2]$,
Note that we need to count the halo pair as a function of the \textit{observed} pairwise velocity between two halos.
We work with the periodic boundary condition for the pair counting.
For a given pairwise velocity $v_{\mathrm{hh}}$, the observed velocity is set by $v_{\mathrm{hh}}+H(z)/(1+z)r_{\parallel}$.
Through these processes, we can have a numerical table for the number of pairs as a function of $\hat{v}_{\alpha}$, 
$r_{p,\beta}$, $M_{i}$, and $M'_{j}$.
Given the table of ${\cal N}(\hat{v}_{\alpha}, r_{p,\beta}, M_{i}, M'_{j})$, we estimate Eq.~(\ref{eq:two_halo_phase_space}) as
\beqa
{\cal F}_{\mathrm{hh}}(\hat{v}_{\alpha}, r_{p,\beta}, M_{i}, M'_{j}) \simeq \
\frac{1}{\Delta \hat{v}}
\frac{1}{2 \pi r_{p, \beta} \Delta r_{p}}
\frac{1}{\Delta M\, \Delta M'}
\frac{1}{L^3_{\mathrm{box}}}
\frac{1}{L^3_{\mathrm{box}}}
{\cal N}(\hat{v}_{\alpha}, r_{p,\beta}, M_{i}, M'_{j}).
\eeqa
Once the numerical table of ${\cal F}_{\mathrm{hh}}$ becomes available, we obtain the relevant two-halo terms to the computation of Eq.~(\ref{eq:hist}) for the ACReS galaxies and LoCuSS clusters as a discrete summation of
\beqa
&&
2 \pi r_{p, \beta} \, \sum_{i,j} \, \Delta M\, \Delta M'\, S(M_{i})\, 
N_{\mathrm{cen}}(M'_{j})\, {\cal F}_{\mathrm{hh}}(\hat{v}_{\alpha}, r_{p, \beta}, M_{i}, M'_{j}) \nonumber \\
&&
+
2 \pi r_{p, \beta} \, \sum_{i,j} \, \Delta M\, \Delta M'\, S(M_{i})\, 
N_{\mathrm{sat}}(M'_{j})\, 
\sum_{p} \Delta \hat{v} \, \tilde{w}_{\mathrm{sat}}(\hat{v}_{p}|M_{i})
{\cal F}_{\mathrm{hh}}(\hat{v}_{\alpha}-\hat{v}_{p}, 
r_{p, \beta}, M_{i}, M'_{j}),
\eeqa
where the former and latter correspond to Eq.~(\ref{eq:cen_cen})
and (\ref{eq:sat_cen}), respectively.
In Eq.~(\ref{eq:sat_cen}), the convolution in velocity appears due to the internal motion of satellites in single halos.
We take into account this convolution by introducing an effective satellite velocity distribution of $\tilde{w}_{\mathrm{sat}}$.
In this paper, we assume that $\tilde{w}_{\mathrm{sat}}$ is a Gaussian with zero mean and the variance of $\sigma^2_{\mathrm{eff}}=\alpha^2_{v} \sigma^2_{\mathrm{DM}}(M, z)$, 
where $M$ is the mass of a host halo,
$\alpha_v$ controls the amplitude of the galaxy velocity dispersion in single clusters,
$z$ is the redshift, and $\sigma_{\mathrm{DM}}$ is given by Eq.~(\ref{eq:sigma_DM}).

\begin{figure*}
\centering
\includegraphics[width=\columnwidth]
{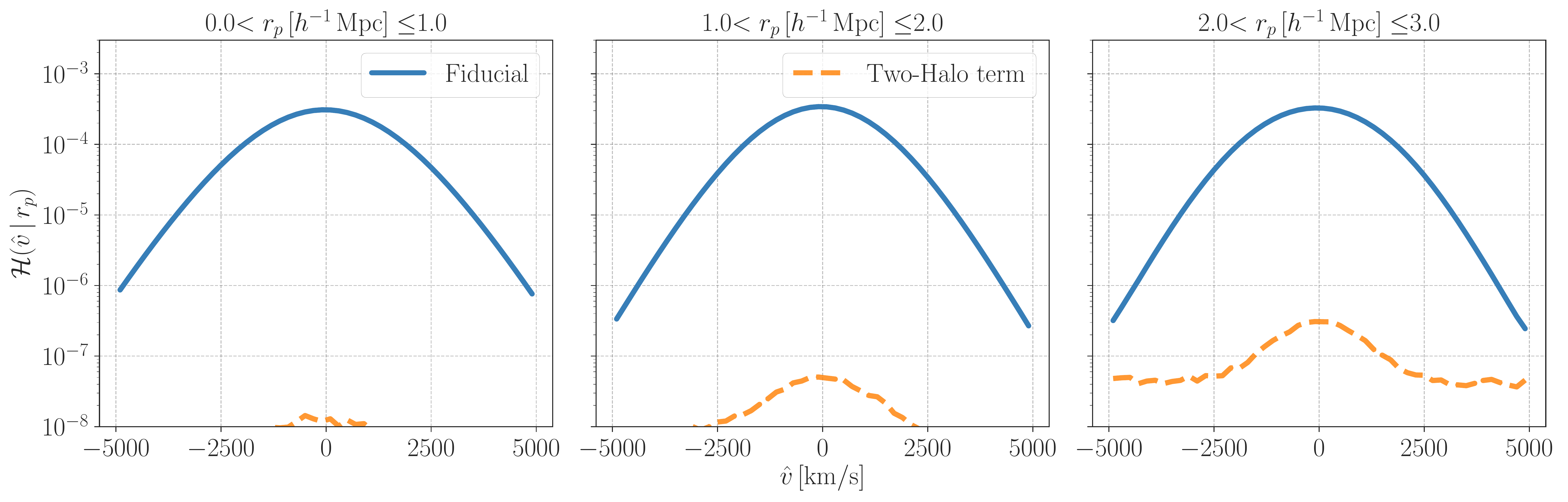}
\caption{
Comparison with one- and two-halo terms in our model of the stacked phase-space density.
The blue lines show the one-halo terms at different radii, while the orange dashed lines represent
the two-halo terms evaluated with the simulation.
}
\label{fig:vPDF_2h}
\end{figure*}

Figure~\ref{fig:vPDF_2h} shows the contribution of the two-halo terms to the stacked phase-space density of ACReS galaxies around the LoCuSS clusters.
In the figure, we adopt the fiducial values of our model parameters as listed in Table~\ref{tb:params} and assume no interlopers at background/foreground of clusters.
We find that the two-halo terms can be safely ignored in our model,
as long as we limit the range of $r_p$ to be less than $\sim 3\, h^{-1}\mathrm{Mpc}$.


\bibliographystyle{mnras}
\bibliography{refs}

\begin{thebibliography}{}
\makeatletter
\relax
\def\mn@urlcharsother{\let\do\@makeother \do\$\do\&\do\#\do\^\do\_\do\%\do\~}
\def\mn@doi{\begingroup\mn@urlcharsother \@ifnextchar [ {\mn@doi@}
  {\mn@doi@[]}}
\def\mn@doi@[#1]#2{\def\@tempa{#1}\ifx\@tempa\@empty \href
  {http://dx.doi.org/#2} {doi:#2}\else \href {http://dx.doi.org/#2} {#1}\fi
  \endgroup}
\def\mn@eprint#1#2{\mn@eprint@#1:#2::\@nil}
\def\mn@eprint@arXiv#1{\href {http://arxiv.org/abs/#1} {{\tt arXiv:#1}}}
\def\mn@eprint@dblp#1{\href {http://dblp.uni-trier.de/rec/bibtex/#1.xml}
  {dblp:#1}}
\def\mn@eprint@#1:#2:#3:#4\@nil{\def\@tempa {#1}\def\@tempb {#2}\def\@tempc
  {#3}\ifx \@tempc \@empty \let \@tempc \@tempb \let \@tempb \@tempa \fi \ifx
  \@tempb \@empty \def\@tempb {arXiv}\fi \@ifundefined
  {mn@eprint@\@tempb}{\@tempb:\@tempc}{\expandafter \expandafter \csname
  mn@eprint@\@tempb\endcsname \expandafter{\@tempc}}}

\bibitem[\protect\citeauthoryear{{Abazajian} et~al.,}{{Abazajian}
  et~al.}{2016}]{2016arXiv161002743A}
{Abazajian} K.~N.,  et~al., 2016, arXiv e-prints, \href
  {https://ui.adsabs.harvard.edu/abs/2016arXiv161002743A} {p. arXiv:1610.02743}

\bibitem[\protect\citeauthoryear{{Abbott} et~al.,}{{Abbott}
  et~al.}{2020}]{2020PhRvD.102b3509A}
{Abbott} T.~M.~C.,  et~al., 2020, \mn@doi [\prd] {10.1103/PhysRevD.102.023509},
  \href {https://ui.adsabs.harvard.edu/abs/2020PhRvD.102b3509A} {102, 023509}

\bibitem[\protect\citeauthoryear{{Adhikari}, {Dalal}, {More}  \&
  {Wetzel}}{{Adhikari} et~al.}{2019}]{2019ApJ...878....9A}
{Adhikari} S.,  {Dalal} N.,  {More} S.,   {Wetzel} A.,  2019, \mn@doi [\apj]
  {10.3847/1538-4357/ab1a39}, \href
  {https://ui.adsabs.harvard.edu/abs/2019ApJ...878....9A} {878, 9}

\bibitem[\protect\citeauthoryear{{Allen}, {Evrard}  \& {Mantz}}{{Allen}
  et~al.}{2011}]{2011ARA&A..49..409A}
{Allen} S.~W.,  {Evrard} A.~E.,   {Mantz} A.~B.,  2011, \mn@doi [\araa]
  {10.1146/annurev-astro-081710-102514}, \href
  {https://ui.adsabs.harvard.edu/abs/2011ARA&A..49..409A} {49, 409}

\bibitem[\protect\citeauthoryear{{Armitage}, {Barnes}, {Kay}, {Bah{\'e}},
  {Dalla Vecchia}, {Crain}  \& {Theuns}}{{Armitage}
  et~al.}{2018}]{2018MNRAS.474.3746A}
{Armitage} T.~J.,  {Barnes} D.~J.,  {Kay} S.~T.,  {Bah{\'e}} Y.~M.,  {Dalla
  Vecchia} C.,  {Crain} R.~A.,   {Theuns} T.,  2018, \mn@doi [\mnras]
  {10.1093/mnras/stx3020}, \href
  {https://ui.adsabs.harvard.edu/abs/2018MNRAS.474.3746A} {474, 3746}

\bibitem[\protect\citeauthoryear{{Arnold}, {Puchwein}  \& {Springel}}{{Arnold}
  et~al.}{2014}]{2014MNRAS.440..833A}
{Arnold} C.,  {Puchwein} E.,   {Springel} V.,  2014, \mn@doi [\mnras]
  {10.1093/mnras/stu332}, \href
  {https://ui.adsabs.harvard.edu/abs/2014MNRAS.440..833A} {440, 833}

\bibitem[\protect\citeauthoryear{{Aung}, {Nagai}, {Rozo}  \&
  {Garc{\'\i}a}}{{Aung} et~al.}{2021}]{2021MNRAS.502.1041A}
{Aung} H.,  {Nagai} D.,  {Rozo} E.,   {Garc{\'\i}a} R.,  2021, \mn@doi [\mnras]
  {10.1093/mnras/staa3994}, \href
  {https://ui.adsabs.harvard.edu/abs/2021MNRAS.502.1041A} {502, 1041}

\bibitem[\protect\citeauthoryear{{Baker} et~al.,}{{Baker}
  et~al.}{2019}]{2019arXiv190803430B}
{Baker} T.,  et~al., 2019, arXiv e-prints, \href
  {https://ui.adsabs.harvard.edu/abs/2019arXiv190803430B} {p. arXiv:1908.03430}

\bibitem[\protect\citeauthoryear{{Barreira}, {Li}, {Jennings}, {Merten},
  {King}, {Baugh}  \& {Pascoli}}{{Barreira} et~al.}{2015}]{2015MNRAS.454.4085B}
{Barreira} A.,  {Li} B.,  {Jennings} E.,  {Merten} J.,  {King} L.,  {Baugh}
  C.~M.,   {Pascoli} S.,  2015, \mn@doi [\mnras] {10.1093/mnras/stv2211}, \href
  {https://ui.adsabs.harvard.edu/abs/2015MNRAS.454.4085B} {454, 4085}

\bibitem[\protect\citeauthoryear{{Behroozi}, {Wechsler}  \& {Wu}}{{Behroozi}
  et~al.}{2013a}]{2013ApJ...762..109B}
{Behroozi} P.~S.,  {Wechsler} R.~H.,   {Wu} H.-Y.,  2013a, \mn@doi [\apj]
  {10.1088/0004-637X/762/2/109}, \href
  {https://ui.adsabs.harvard.edu/abs/2013ApJ...762..109B} {762, 109}

\bibitem[\protect\citeauthoryear{{Behroozi}, {Wechsler}  \&
  {Conroy}}{{Behroozi} et~al.}{2013b}]{2013ApJ...770...57B}
{Behroozi} P.~S.,  {Wechsler} R.~H.,   {Conroy} C.,  2013b, \mn@doi [\apj]
  {10.1088/0004-637X/770/1/57}, \href
  {https://ui.adsabs.harvard.edu/abs/2013ApJ...770...57B} {770, 57}

\bibitem[\protect\citeauthoryear{{Bianchi}, {Percival}  \& {Bel}}{{Bianchi}
  et~al.}{2016}]{2016MNRAS.463.3783B}
{Bianchi} D.,  {Percival} W.~J.,   {Bel} J.,  2016, \mn@doi [\mnras]
  {10.1093/mnras/stw2243}, \href
  {https://ui.adsabs.harvard.edu/abs/2016MNRAS.463.3783B} {463, 3783}

\bibitem[\protect\citeauthoryear{{Bianconi}, {Buscicchio}, {Smith}, {McGee},
  {Haines}, {Finoguenov}  \& {Babul}}{{Bianconi}
  et~al.}{2020}]{2020arXiv201005920B}
{Bianconi} M.,  {Buscicchio} R.,  {Smith} G.~P.,  {McGee} S.~L.,  {Haines}
  C.~P.,  {Finoguenov} A.,   {Babul} A.,  2020, arXiv e-prints, \href
  {https://ui.adsabs.harvard.edu/abs/2020arXiv201005920B} {p. arXiv:2010.05920}

\bibitem[\protect\citeauthoryear{{Biviano}, {Murante}, {Borgani}, {Diaferio},
  {Dolag}  \& {Girardi}}{{Biviano} et~al.}{2006}]{2006A&A...456...23B}
{Biviano} A.,  {Murante} G.,  {Borgani} S.,  {Diaferio} A.,  {Dolag} K.,
  {Girardi} M.,  2006, \mn@doi [\aap] {10.1051/0004-6361:20064918}, \href
  {https://ui.adsabs.harvard.edu/abs/2006A&A...456...23B} {456, 23}

\bibitem[\protect\citeauthoryear{{Biviano} et~al.,}{{Biviano}
  et~al.}{2013}]{2013A&A...558A...1B}
{Biviano} A.,  et~al., 2013, \mn@doi [\aap] {10.1051/0004-6361/201321955},
  \href {https://ui.adsabs.harvard.edu/abs/2013A&A...558A...1B} {558, A1}

\bibitem[\protect\citeauthoryear{{Bocquet} et~al.,}{{Bocquet}
  et~al.}{2019}]{2019ApJ...878...55B}
{Bocquet} S.,  et~al., 2019, \mn@doi [\apj] {10.3847/1538-4357/ab1f10}, \href
  {https://ui.adsabs.harvard.edu/abs/2019ApJ...878...55B} {878, 55}

\bibitem[\protect\citeauthoryear{{B{\"o}hringer} et~al.,}{{B{\"o}hringer}
  et~al.}{2004}]{2004A&A...425..367B}
{B{\"o}hringer} H.,  et~al., 2004, \mn@doi [\aap] {10.1051/0004-6361:20034484},
  \href {https://ui.adsabs.harvard.edu/abs/2004A&A...425..367B} {425, 367}

\bibitem[\protect\citeauthoryear{{Cappi}}{{Cappi}}{1995}]{1995A&A...301....6C}
{Cappi} A.,  1995, \aap, \href
  {https://ui.adsabs.harvard.edu/abs/1995A&A...301....6C} {301, 6}

\bibitem[\protect\citeauthoryear{{Carlberg} et~al.,}{{Carlberg}
  et~al.}{1997}]{1997ApJ...485L..13C}
{Carlberg} R.~G.,  et~al., 1997, \mn@doi [\apjl] {10.1086/310801}, \href
  {https://ui.adsabs.harvard.edu/abs/1997ApJ...485L..13C} {485, L13}

\bibitem[\protect\citeauthoryear{{Chang} et~al.,}{{Chang}
  et~al.}{2018}]{2018ApJ...864...83C}
{Chang} C.,  et~al., 2018, \mn@doi [\apj] {10.3847/1538-4357/aad5e7}, \href
  {https://ui.adsabs.harvard.edu/abs/2018ApJ...864...83C} {864, 83}

\bibitem[\protect\citeauthoryear{{Cole} \& {Lacey}}{{Cole} \&
  {Lacey}}{1996}]{1996MNRAS.281..716C}
{Cole} S.,  {Lacey} C.,  1996, \mn@doi [\mnras] {10.1093/mnras/281.2.716},
  \href {https://ui.adsabs.harvard.edu/abs/1996MNRAS.281..716C} {281, 716}

\bibitem[\protect\citeauthoryear{{Cooray} \& {Sheth}}{{Cooray} \&
  {Sheth}}{2002}]{2002PhR...372....1C}
{Cooray} A.,  {Sheth} R.,  2002, \mn@doi [\physrep]
  {10.1016/S0370-1573(02)00276-4}, \href
  {https://ui.adsabs.harvard.edu/abs/2002PhR...372....1C} {372, 1}

\bibitem[\protect\citeauthoryear{{Cuesta-Lazaro}, {Li}, {Eggemeier}, {Zarrouk},
  {Baugh}, {Nishimichi}  \& {Takada}}{{Cuesta-Lazaro}
  et~al.}{2020}]{2020MNRAS.498.1175C}
{Cuesta-Lazaro} C.,  {Li} B.,  {Eggemeier} A.,  {Zarrouk} P.,  {Baugh} C.~M.,
  {Nishimichi} T.,   {Takada} M.,  2020, \mn@doi [\mnras]
  {10.1093/mnras/staa2249}, \href
  {https://ui.adsabs.harvard.edu/abs/2020MNRAS.498.1175C} {498, 1175}

\bibitem[\protect\citeauthoryear{{Damjanov}, {Zahid}, {Geller}, {Fabricant}  \&
  {Hwang}}{{Damjanov} et~al.}{2018}]{2018ApJS..234...21D}
{Damjanov} I.,  {Zahid} H.~J.,  {Geller} M.~J.,  {Fabricant} D.~G.,   {Hwang}
  H.~S.,  2018, \mn@doi [\apjs] {10.3847/1538-4365/aaa01c}, \href
  {https://ui.adsabs.harvard.edu/abs/2018ApJS..234...21D} {234, 21}

\bibitem[\protect\citeauthoryear{{Diaferio} \& {Geller}}{{Diaferio} \&
  {Geller}}{1997}]{1997ApJ...481..633D}
{Diaferio} A.,  {Geller} M.~J.,  1997, \mn@doi [\apj] {10.1086/304075}, \href
  {https://ui.adsabs.harvard.edu/abs/1997ApJ...481..633D} {481, 633}

\bibitem[\protect\citeauthoryear{{Diemer} \& {Kravtsov}}{{Diemer} \&
  {Kravtsov}}{2015}]{2015ApJ...799..108D}
{Diemer} B.,  {Kravtsov} A.~V.,  2015, \mn@doi [\apj]
  {10.1088/0004-637X/799/1/108}, \href
  {https://ui.adsabs.harvard.edu/abs/2015ApJ...799..108D} {799, 108}

\bibitem[\protect\citeauthoryear{{Dvali}, {Gabadadze}  \& {Porrati}}{{Dvali}
  et~al.}{2000}]{2000PhLB..485..208D}
{Dvali} G.,  {Gabadadze} G.,   {Porrati} M.,  2000, \mn@doi [Physics Letters B]
  {10.1016/S0370-2693(00)00669-9}, \href
  {https://ui.adsabs.harvard.edu/abs/2000PhLB..485..208D} {485, 208}

\bibitem[\protect\citeauthoryear{{Ebeling}, {Edge}, {Bohringer}, {Allen},
  {Crawford}, {Fabian}, {Voges}  \& {Huchra}}{{Ebeling}
  et~al.}{1998}]{1998MNRAS.301..881E}
{Ebeling} H.,  {Edge} A.~C.,  {Bohringer} H.,  {Allen} S.~W.,  {Crawford}
  C.~S.,  {Fabian} A.~C.,  {Voges} W.,   {Huchra} J.~P.,  1998, \mn@doi
  [\mnras] {10.1046/j.1365-8711.1998.01949.x}, \href
  {https://ui.adsabs.harvard.edu/abs/1998MNRAS.301..881E} {301, 881}

\bibitem[\protect\citeauthoryear{{Ebeling}, {Edge}, {Allen}, {Crawford},
  {Fabian}  \& {Huchra}}{{Ebeling} et~al.}{2000}]{2000MNRAS.318..333E}
{Ebeling} H.,  {Edge} A.~C.,  {Allen} S.~W.,  {Crawford} C.~S.,  {Fabian}
  A.~C.,   {Huchra} J.~P.,  2000, \mn@doi [\mnras]
  {10.1046/j.1365-8711.2000.03549.x}, \href
  {https://ui.adsabs.harvard.edu/abs/2000MNRAS.318..333E} {318, 333}

\bibitem[\protect\citeauthoryear{{Evrard} et~al.,}{{Evrard}
  et~al.}{2008}]{2008ApJ...672..122E}
{Evrard} A.~E.,  et~al., 2008, \mn@doi [\apj] {10.1086/521616}, \href
  {https://ui.adsabs.harvard.edu/abs/2008ApJ...672..122E} {672, 122}

\bibitem[\protect\citeauthoryear{{Fairbairn} \& {Goobar}}{{Fairbairn} \&
  {Goobar}}{2006}]{2006PhLB..642..432F}
{Fairbairn} M.,  {Goobar} A.,  2006, \mn@doi [Physics Letters B]
  {10.1016/j.physletb.2006.07.048}, \href
  {https://ui.adsabs.harvard.edu/abs/2006PhLB..642..432F} {642, 432}

\bibitem[\protect\citeauthoryear{{Fang}, {Wang}, {Hu}, {Haiman}, {Hui}  \&
  {May}}{{Fang} et~al.}{2008}]{2008PhRvD..78j3509F}
{Fang} W.,  {Wang} S.,  {Hu} W.,  {Haiman} Z.,  {Hui} L.,   {May} M.,  2008,
  \mn@doi [\prd] {10.1103/PhysRevD.78.103509}, \href
  {https://ui.adsabs.harvard.edu/abs/2008PhRvD..78j3509F} {78, 103509}

\bibitem[\protect\citeauthoryear{{Fert{\'e}}, {Kirk}, {Liddle}  \&
  {Zuntz}}{{Fert{\'e}} et~al.}{2019}]{2019PhRvD..99h3512F}
{Fert{\'e}} A.,  {Kirk} D.,  {Liddle} A.~R.,   {Zuntz} J.,  2019, \mn@doi
  [\prd] {10.1103/PhysRevD.99.083512}, \href
  {https://ui.adsabs.harvard.edu/abs/2019PhRvD..99h3512F} {99, 083512}

\bibitem[\protect\citeauthoryear{{Foreman-Mackey}, {Hogg}, {Lang}  \&
  {Goodman}}{{Foreman-Mackey} et~al.}{2013}]{2013PASP..125..306F}
{Foreman-Mackey} D.,  {Hogg} D.~W.,  {Lang} D.,   {Goodman} J.,  2013, \mn@doi
  [\pasp] {10.1086/670067}, \href
  {https://ui.adsabs.harvard.edu/abs/2013PASP..125..306F} {125, 306}

\bibitem[\protect\citeauthoryear{{Garcia-Quintero}, {Ishak}  \&
  {Ning}}{{Garcia-Quintero} et~al.}{2020}]{2020JCAP...12..018G}
{Garcia-Quintero} C.,  {Ishak} M.,   {Ning} O.,  2020, \mn@doi [\jcap]
  {10.1088/1475-7516/2020/12/018}, \href
  {https://ui.adsabs.harvard.edu/abs/2020JCAP...12..018G} {2020, 018}

\bibitem[\protect\citeauthoryear{{Geller}, {Hwang}, {Diaferio}, {Kurtz}, {Coe}
  \& {Rines}}{{Geller} et~al.}{2014}]{2014ApJ...783...52G}
{Geller} M.~J.,  {Hwang} H.~S.,  {Diaferio} A.,  {Kurtz} M.~J.,  {Coe} D.,
  {Rines} K.~J.,  2014, \mn@doi [\apj] {10.1088/0004-637X/783/1/52}, \href
  {https://ui.adsabs.harvard.edu/abs/2014ApJ...783...52G} {783, 52}

\bibitem[\protect\citeauthoryear{{Haines} et~al.,}{{Haines}
  et~al.}{2013}]{2013ApJ...775..126H}
{Haines} C.~P.,  et~al., 2013, \mn@doi [\apj] {10.1088/0004-637X/775/2/126},
  \href {https://ui.adsabs.harvard.edu/abs/2013ApJ...775..126H} {775, 126}

\bibitem[\protect\citeauthoryear{{Haines} et~al.,}{{Haines}
  et~al.}{2015}]{2015ApJ...806..101H}
{Haines} C.~P.,  et~al., 2015, \mn@doi [\apj] {10.1088/0004-637X/806/1/101},
  \href {https://ui.adsabs.harvard.edu/abs/2015ApJ...806..101H} {806, 101}

\bibitem[\protect\citeauthoryear{{Hamabata}, {Oogi}, {Oguri}, {Nishimichi}  \&
  {Nagashima}}{{Hamabata} et~al.}{2019a}]{2019MNRAS.488.4117H}
{Hamabata} A.,  {Oogi} T.,  {Oguri} M.,  {Nishimichi} T.,   {Nagashima} M.,
  2019a, \mn@doi [\mnras] {10.1093/mnras/stz1991}, \href
  {https://ui.adsabs.harvard.edu/abs/2019MNRAS.488.4117H} {488, 4117}

\bibitem[\protect\citeauthoryear{{Hamabata}, {Oguri}  \&
  {Nishimichi}}{{Hamabata} et~al.}{2019b}]{2019MNRAS.489.1344H}
{Hamabata} A.,  {Oguri} M.,   {Nishimichi} T.,  2019b, \mn@doi [\mnras]
  {10.1093/mnras/stz2227}, \href
  {https://ui.adsabs.harvard.edu/abs/2019MNRAS.489.1344H} {489, 1344}

\bibitem[\protect\citeauthoryear{{Hansen} \& {Moore}}{{Hansen} \&
  {Moore}}{2006}]{2006NewA...11..333H}
{Hansen} S.~H.,  {Moore} B.,  2006, \mn@doi [\na]
  {10.1016/j.newast.2005.09.001}, \href
  {https://ui.adsabs.harvard.edu/abs/2006NewA...11..333H} {11, 333}

\bibitem[\protect\citeauthoryear{{Hellwing}, {Barreira}, {Frenk}, {Li}  \&
  {Cole}}{{Hellwing} et~al.}{2014}]{2014PhRvL.112v1102H}
{Hellwing} W.~A.,  {Barreira} A.,  {Frenk} C.~S.,  {Li} B.,   {Cole} S.,  2014,
  \mn@doi [\prl] {10.1103/PhysRevLett.112.221102}, \href
  {https://ui.adsabs.harvard.edu/abs/2014PhRvL.112v1102H} {112, 221102}

\bibitem[\protect\citeauthoryear{Hu \& Kravtsov}{Hu \&
  Kravtsov}{2003}]{Hu:2002we}
Hu W.,  Kravtsov A.~V.,  2003, \mn@doi [Astrophys. J.] {10.1086/345846}, 584,
  702

\bibitem[\protect\citeauthoryear{{Hu} \& {Sawicki}}{{Hu} \&
  {Sawicki}}{2007}]{2007PhRvD..76f4004H}
{Hu} W.,  {Sawicki} I.,  2007, \mn@doi [\prd] {10.1103/PhysRevD.76.064004},
  \href {https://ui.adsabs.harvard.edu/abs/2007PhRvD..76f4004H} {76, 064004}

\bibitem[\protect\citeauthoryear{{Ishikawa} et~al.,}{{Ishikawa}
  et~al.}{2020}]{2020ApJ...904..128I}
{Ishikawa} S.,  et~al., 2020, \mn@doi [\apj] {10.3847/1538-4357/abbd95}, \href
  {https://ui.adsabs.harvard.edu/abs/2020ApJ...904..128I} {904, 128}

\bibitem[\protect\citeauthoryear{{Ishiyama}, {Enoki}, {Kobayashi}, {Makiya},
  {Nagashima}  \& {Oogi}}{{Ishiyama} et~al.}{2015}]{2015PASJ...67...61I}
{Ishiyama} T.,  {Enoki} M.,  {Kobayashi} M. A.~R.,  {Makiya} R.,  {Nagashima}
  M.,   {Oogi} T.,  2015, \mn@doi [\pasj] {10.1093/pasj/psv021}, \href
  {https://ui.adsabs.harvard.edu/abs/2015PASJ...67...61I} {67, 61}

\bibitem[\protect\citeauthoryear{{Kaiser}}{{Kaiser}}{2013}]{2013MNRAS.435.1278K}
{Kaiser} N.,  2013, \mn@doi [\mnras] {10.1093/mnras/stt1370}, \href
  {https://ui.adsabs.harvard.edu/abs/2013MNRAS.435.1278K} {435, 1278}

\bibitem[\protect\citeauthoryear{{Kim} \& {Croft}}{{Kim} \&
  {Croft}}{2004}]{2004ApJ...607..164K}
{Kim} Y.-R.,  {Croft} R. A.~C.,  2004, \mn@doi [\apj] {10.1086/383218}, \href
  {https://ui.adsabs.harvard.edu/abs/2004ApJ...607..164K} {607, 164}

\bibitem[\protect\citeauthoryear{{Kobayashi}, {Watanabe}  \&
  {Yamauchi}}{{Kobayashi} et~al.}{2015}]{2015PhRvD..91f4013K}
{Kobayashi} T.,  {Watanabe} Y.,   {Yamauchi} D.,  2015, \mn@doi [\prd]
  {10.1103/PhysRevD.91.064013}, \href
  {https://ui.adsabs.harvard.edu/abs/2015PhRvD..91f4013K} {91, 064013}

\bibitem[\protect\citeauthoryear{{Kuruvilla} \& {Porciani}}{{Kuruvilla} \&
  {Porciani}}{2018}]{2018MNRAS.479.2256K}
{Kuruvilla} J.,  {Porciani} C.,  2018, \mn@doi [\mnras]
  {10.1093/mnras/sty1654}, \href
  {https://ui.adsabs.harvard.edu/abs/2018MNRAS.479.2256K} {479, 2256}

\bibitem[\protect\citeauthoryear{{Lam}, {Nishimichi}, {Schmidt}  \&
  {Takada}}{{Lam} et~al.}{2012}]{2012PhRvL.109e1301L}
{Lam} T.~Y.,  {Nishimichi} T.,  {Schmidt} F.,   {Takada} M.,  2012, \mn@doi
  [\prl] {10.1103/PhysRevLett.109.051301}, \href
  {https://ui.adsabs.harvard.edu/abs/2012PhRvL.109e1301L} {109, 051301}

\bibitem[\protect\citeauthoryear{{Lam}, {Schmidt}, {Nishimichi}  \&
  {Takada}}{{Lam} et~al.}{2013}]{2013PhRvD..88b3012L}
{Lam} T.~Y.,  {Schmidt} F.,  {Nishimichi} T.,   {Takada} M.,  2013, \mn@doi
  [\prd] {10.1103/PhysRevD.88.023012}, \href
  {https://ui.adsabs.harvard.edu/abs/2013PhRvD..88b3012L} {88, 023012}

\bibitem[\protect\citeauthoryear{{Lau}, {Nagai}  \& {Kravtsov}}{{Lau}
  et~al.}{2010}]{2010ApJ...708.1419L}
{Lau} E.~T.,  {Nagai} D.,   {Kravtsov} A.~V.,  2010, \mn@doi [\apj]
  {10.1088/0004-637X/708/2/1419}, \href
  {https://ui.adsabs.harvard.edu/abs/2010ApJ...708.1419L} {708, 1419}

\bibitem[\protect\citeauthoryear{{{\L}okas} \& {Mamon}}{{{\L}okas} \&
  {Mamon}}{2001}]{2001MNRAS.321..155L}
{{\L}okas} E.~L.,  {Mamon} G.~A.,  2001, \mn@doi [\mnras]
  {10.1046/j.1365-8711.2001.04007.x}, \href
  {https://ui.adsabs.harvard.edu/abs/2001MNRAS.321..155L} {321, 155}

\bibitem[\protect\citeauthoryear{{Lombriser}, {Hu}, {Fang}  \&
  {Seljak}}{{Lombriser} et~al.}{2009}]{2009PhRvD..80f3536L}
{Lombriser} L.,  {Hu} W.,  {Fang} W.,   {Seljak} U.,  2009, \mn@doi [\prd]
  {10.1103/PhysRevD.80.063536}, \href
  {https://ui.adsabs.harvard.edu/abs/2009PhRvD..80f3536L} {80, 063536}

\bibitem[\protect\citeauthoryear{{Maartens} \& {Majerotto}}{{Maartens} \&
  {Majerotto}}{2006}]{2006PhRvD..74b3004M}
{Maartens} R.,  {Majerotto} E.,  2006, \mn@doi [\prd]
  {10.1103/PhysRevD.74.023004}, \href
  {https://ui.adsabs.harvard.edu/abs/2006PhRvD..74b3004M} {74, 023004}

\bibitem[\protect\citeauthoryear{{Mamon}, {Biviano}  \& {Bou{\'e}}}{{Mamon}
  et~al.}{2013}]{2013MNRAS.429.3079M}
{Mamon} G.~A.,  {Biviano} A.,   {Bou{\'e}} G.,  2013, \mn@doi [\mnras]
  {10.1093/mnras/sts565}, \href
  {https://ui.adsabs.harvard.edu/abs/2013MNRAS.429.3079M} {429, 3079}

\bibitem[\protect\citeauthoryear{{Mantz}, {Allen}, {Morris}, {Rapetti},
  {Applegate}, {Kelly}, {von der Linden}  \& {Schmidt}}{{Mantz}
  et~al.}{2014}]{2014MNRAS.440.2077M}
{Mantz} A.~B.,  {Allen} S.~W.,  {Morris} R.~G.,  {Rapetti} D.~A.,  {Applegate}
  D.~E.,  {Kelly} P.~L.,  {von der Linden} A.,   {Schmidt} R.~W.,  2014,
  \mn@doi [\mnras] {10.1093/mnras/stu368}, \href
  {https://ui.adsabs.harvard.edu/abs/2014MNRAS.440.2077M} {440, 2077}

\bibitem[\protect\citeauthoryear{{Martel}, {Robichaud}  \& {Barai}}{{Martel}
  et~al.}{2014}]{2014ApJ...786...79M}
{Martel} H.,  {Robichaud} F.,   {Barai} P.,  2014, \mn@doi [\apj]
  {10.1088/0004-637X/786/2/79}, \href
  {https://ui.adsabs.harvard.edu/abs/2014ApJ...786...79M} {786, 79}

\bibitem[\protect\citeauthoryear{{Mercurio}, {Girardi}, {Boschin}, {Merluzzi}
  \& {Busarello}}{{Mercurio} et~al.}{2003}]{2003A&A...397..431M}
{Mercurio} A.,  {Girardi} M.,  {Boschin} W.,  {Merluzzi} P.,   {Busarello} G.,
  2003, \mn@doi [\aap] {10.1051/0004-6361:20021495}, \href
  {https://ui.adsabs.harvard.edu/abs/2003A&A...397..431M} {397, 431}

\bibitem[\protect\citeauthoryear{{Merloni} et~al.,}{{Merloni}
  et~al.}{2012}]{2012arXiv1209.3114M}
{Merloni} A.,  et~al., 2012, arXiv e-prints, \href
  {https://ui.adsabs.harvard.edu/abs/2012arXiv1209.3114M} {p. arXiv:1209.3114}

\bibitem[\protect\citeauthoryear{{Mitchell}, {He}, {Arnold}  \&
  {Li}}{{Mitchell} et~al.}{2018}]{2018MNRAS.477.1133M}
{Mitchell} M.~A.,  {He} J.-h.,  {Arnold} C.,   {Li} B.,  2018, \mn@doi [\mnras]
  {10.1093/mnras/sty636}, \href
  {https://ui.adsabs.harvard.edu/abs/2018MNRAS.477.1133M} {477, 1133}

\bibitem[\protect\citeauthoryear{{More}, {van den Bosch}  \& {Cacciato}}{{More}
  et~al.}{2009}]{2009MNRAS.392..917M}
{More} S.,  {van den Bosch} F.~C.,   {Cacciato} M.,  2009, \mn@doi [\mnras]
  {10.1111/j.1365-2966.2008.14114.x}, \href
  {https://ui.adsabs.harvard.edu/abs/2009MNRAS.392..917M} {392, 917}

\bibitem[\protect\citeauthoryear{{More}, {Diemer}  \& {Kravtsov}}{{More}
  et~al.}{2015}]{2015ApJ...810...36M}
{More} S.,  {Diemer} B.,   {Kravtsov} A.~V.,  2015, \mn@doi [\apj]
  {10.1088/0004-637X/810/1/36}, \href
  {https://ui.adsabs.harvard.edu/abs/2015ApJ...810...36M} {810, 36}

\bibitem[\protect\citeauthoryear{{More} et~al.,}{{More}
  et~al.}{2016}]{2016ApJ...825...39M}
{More} S.,  et~al., 2016, \mn@doi [\apj] {10.3847/0004-637X/825/1/39}, \href
  {https://ui.adsabs.harvard.edu/abs/2016ApJ...825...39M} {825, 39}

\bibitem[\protect\citeauthoryear{{Munari}, {Biviano}, {Borgani}, {Murante}  \&
  {Fabjan}}{{Munari} et~al.}{2013}]{2013MNRAS.430.2638M}
{Munari} E.,  {Biviano} A.,  {Borgani} S.,  {Murante} G.,   {Fabjan} D.,  2013,
  \mn@doi [\mnras] {10.1093/mnras/stt049}, \href
  {https://ui.adsabs.harvard.edu/abs/2013MNRAS.430.2638M} {430, 2638}

\bibitem[\protect\citeauthoryear{{Murata}, {Sunayama}, {Oguri}, {More},
  {Nishizawa}, {Nishimichi}  \& {Osato}}{{Murata}
  et~al.}{2020}]{2020PASJ...72...64M}
{Murata} R.,  {Sunayama} T.,  {Oguri} M.,  {More} S.,  {Nishizawa} A.~J.,
  {Nishimichi} T.,   {Osato} K.,  2020, \mn@doi [\pasj] {10.1093/pasj/psaa041},
  \href {https://ui.adsabs.harvard.edu/abs/2020PASJ...72...64M} {72, 64}

\bibitem[\protect\citeauthoryear{{Navarro}, {Frenk}  \& {White}}{{Navarro}
  et~al.}{1996}]{1996ApJ...462..563N}
{Navarro} J.~F.,  {Frenk} C.~S.,   {White} S. D.~M.,  1996, \mn@doi [\apj]
  {10.1086/177173}, \href
  {https://ui.adsabs.harvard.edu/abs/1996ApJ...462..563N} {462, 563}

\bibitem[\protect\citeauthoryear{{Navarro}, {Frenk}  \& {White}}{{Navarro}
  et~al.}{1997}]{1997ApJ...490..493N}
{Navarro} J.~F.,  {Frenk} C.~S.,   {White} S. D.~M.,  1997, \mn@doi [\apj]
  {10.1086/304888}, \href
  {https://ui.adsabs.harvard.edu/abs/1997ApJ...490..493N} {490, 493}

\bibitem[\protect\citeauthoryear{{Okabe} \& {Smith}}{{Okabe} \&
  {Smith}}{2016}]{2016MNRAS.461.3794O}
{Okabe} N.,  {Smith} G.~P.,  2016, \mn@doi [\mnras] {10.1093/mnras/stw1539},
  \href {https://ui.adsabs.harvard.edu/abs/2016MNRAS.461.3794O} {461, 3794}

\bibitem[\protect\citeauthoryear{{Okabe}, {Takada}, {Umetsu}, {Futamase}  \&
  {Smith}}{{Okabe} et~al.}{2010}]{2010PASJ...62..811O}
{Okabe} N.,  {Takada} M.,  {Umetsu} K.,  {Futamase} T.,   {Smith} G.~P.,  2010,
  \mn@doi [\pasj] {10.1093/pasj/62.3.811}, \href
  {https://ui.adsabs.harvard.edu/abs/2010PASJ...62..811O} {62, 811}

\bibitem[\protect\citeauthoryear{{Owers}, {Nulsen}  \& {Couch}}{{Owers}
  et~al.}{2011}]{2011ApJ...741..122O}
{Owers} M.~S.,  {Nulsen} P. E.~J.,   {Couch} W.~J.,  2011, \mn@doi [\apj]
  {10.1088/0004-637X/741/2/122}, \href
  {https://ui.adsabs.harvard.edu/abs/2011ApJ...741..122O} {741, 122}

\bibitem[\protect\citeauthoryear{{Planck Collaboration} et~al.,}{{Planck
  Collaboration} et~al.}{2016a}]{2016A&A...594A..13P}
{Planck Collaboration} et~al., 2016a, \mn@doi [\aap]
  {10.1051/0004-6361/201525830}, \href
  {https://ui.adsabs.harvard.edu/abs/2016A&A...594A..13P} {594, A13}

\bibitem[\protect\citeauthoryear{{Planck Collaboration} et~al.,}{{Planck
  Collaboration} et~al.}{2016b}]{2016A&A...594A..24P}
{Planck Collaboration} et~al., 2016b, \mn@doi [\aap]
  {10.1051/0004-6361/201525833}, \href
  {https://ui.adsabs.harvard.edu/abs/2016A&A...594A..24P} {594, A24}

\bibitem[\protect\citeauthoryear{{Raccanelli} et~al.,}{{Raccanelli}
  et~al.}{2013}]{2013MNRAS.436...89R}
{Raccanelli} A.,  et~al., 2013, \mn@doi [\mnras] {10.1093/mnras/stt1517}, \href
  {https://ui.adsabs.harvard.edu/abs/2013MNRAS.436...89R} {436, 89}

\bibitem[\protect\citeauthoryear{{Rines}, {Geller}, {Diaferio}  \&
  {Kurtz}}{{Rines} et~al.}{2013}]{2013ApJ...767...15R}
{Rines} K.,  {Geller} M.~J.,  {Diaferio} A.,   {Kurtz} M.~J.,  2013, \mn@doi
  [\apj] {10.1088/0004-637X/767/1/15}, \href
  {https://ui.adsabs.harvard.edu/abs/2013ApJ...767...15R} {767, 15}

\bibitem[\protect\citeauthoryear{{Rozo} et~al.,}{{Rozo}
  et~al.}{2010}]{2010ApJ...708..645R}
{Rozo} E.,  et~al., 2010, \mn@doi [\apj] {10.1088/0004-637X/708/1/645}, \href
  {https://ui.adsabs.harvard.edu/abs/2010ApJ...708..645R} {708, 645}

\bibitem[\protect\citeauthoryear{{Saro}, {Mohr}, {Bazin}  \& {Dolag}}{{Saro}
  et~al.}{2013}]{2013ApJ...772...47S}
{Saro} A.,  {Mohr} J.~J.,  {Bazin} G.,   {Dolag} K.,  2013, \mn@doi [\apj]
  {10.1088/0004-637X/772/1/47}, \href
  {https://ui.adsabs.harvard.edu/abs/2013ApJ...772...47S} {772, 47}

\bibitem[\protect\citeauthoryear{{Schechter}}{{Schechter}}{1976}]{1976ApJ...203..297S}
{Schechter} P.,  1976, \mn@doi [\apj] {10.1086/154079}, \href
  {https://ui.adsabs.harvard.edu/abs/1976ApJ...203..297S} {203, 297}

\bibitem[\protect\citeauthoryear{{Schmidt}}{{Schmidt}}{2009}]{2009PhRvD..80d3001S}
{Schmidt} F.,  2009, \mn@doi [\prd] {10.1103/PhysRevD.80.043001}, \href
  {https://ui.adsabs.harvard.edu/abs/2009PhRvD..80d3001S} {80, 043001}

\bibitem[\protect\citeauthoryear{{Schmidt}}{{Schmidt}}{2010}]{2010PhRvD..81j3002S}
{Schmidt} F.,  2010, \mn@doi [\prd] {10.1103/PhysRevD.81.103002}, \href
  {https://ui.adsabs.harvard.edu/abs/2010PhRvD..81j3002S} {81, 103002}

\bibitem[\protect\citeauthoryear{{Schmidt}, {Hu}  \& {Lima}}{{Schmidt}
  et~al.}{2010}]{2010PhRvD..81f3005S}
{Schmidt} F.,  {Hu} W.,   {Lima} M.,  2010, \mn@doi [\prd]
  {10.1103/PhysRevD.81.063005}, \href
  {https://ui.adsabs.harvard.edu/abs/2010PhRvD..81f3005S} {81, 063005}

\bibitem[\protect\citeauthoryear{{Shirasaki}, {Huff}, {Markovic}  \&
  {Rhodes}}{{Shirasaki} et~al.}{2021}]{2021ApJ...907...38S}
{Shirasaki} M.,  {Huff} E.~M.,  {Markovic} K.,   {Rhodes} J.~D.,  2021, \mn@doi
  [\apj] {10.3847/1538-4357/abcc68}, \href
  {https://ui.adsabs.harvard.edu/abs/2021ApJ...907...38S} {907, 38}

\bibitem[\protect\citeauthoryear{{Simpson} et~al.,}{{Simpson}
  et~al.}{2013}]{2013MNRAS.429.2249S}
{Simpson} F.,  et~al., 2013, \mn@doi [\mnras] {10.1093/mnras/sts493}, \href
  {https://ui.adsabs.harvard.edu/abs/2013MNRAS.429.2249S} {429, 2249}

\bibitem[\protect\citeauthoryear{{Skibba}, {van den Bosch}, {Yang}, {More},
  {Mo}  \& {Fontanot}}{{Skibba} et~al.}{2011}]{2011MNRAS.410..417S}
{Skibba} R.~A.,  {van den Bosch} F.~C.,  {Yang} X.,  {More} S.,  {Mo} H.,
  {Fontanot} F.,  2011, \mn@doi [\mnras] {10.1111/j.1365-2966.2010.17452.x},
  \href {https://ui.adsabs.harvard.edu/abs/2011MNRAS.410..417S} {410, 417}

\bibitem[\protect\citeauthoryear{{Smith} et~al.,}{{Smith}
  et~al.}{2016}]{2016MNRAS.456L..74S}
{Smith} G.~P.,  et~al., 2016, \mn@doi [\mnras] {10.1093/mnrasl/slv175}, \href
  {https://ui.adsabs.harvard.edu/abs/2016MNRAS.456L..74S} {456, L74}

\bibitem[\protect\citeauthoryear{{Stark}, {Miller}  \& {Halenka}}{{Stark}
  et~al.}{2019}]{2019ApJ...874...33S}
{Stark} A.,  {Miller} C.~J.,   {Halenka} V.,  2019, \mn@doi [\apj]
  {10.3847/1538-4357/ab06fa}, \href
  {https://ui.adsabs.harvard.edu/abs/2019ApJ...874...33S} {874, 33}

\bibitem[\protect\citeauthoryear{{Terukina}, {Lombriser}, {Yamamoto}, {Bacon},
  {Koyama}  \& {Nichol}}{{Terukina} et~al.}{2014}]{2014JCAP...04..013T}
{Terukina} A.,  {Lombriser} L.,  {Yamamoto} K.,  {Bacon} D.,  {Koyama} K.,
  {Nichol} R.~C.,  2014, \mn@doi [\jcap] {10.1088/1475-7516/2014/04/013}, \href
  {https://ui.adsabs.harvard.edu/abs/2014JCAP...04..013T} {2014, 013}

\bibitem[\protect\citeauthoryear{{Terukina}, {Yamamoto}, {Okabe}, {Matsushita}
  \& {Sasaki}}{{Terukina} et~al.}{2015}]{2015JCAP...10..064T}
{Terukina} A.,  {Yamamoto} K.,  {Okabe} N.,  {Matsushita} K.,   {Sasaki} T.,
  2015, \mn@doi [\jcap] {10.1088/1475-7516/2015/10/064}, \href
  {https://ui.adsabs.harvard.edu/abs/2015JCAP...10..064T} {2015, 064}

\bibitem[\protect\citeauthoryear{{Tinker}}{{Tinker}}{2007}]{2007MNRAS.374..477T}
{Tinker} J.~L.,  2007, \mn@doi [\mnras] {10.1111/j.1365-2966.2006.11157.x},
  \href {https://ui.adsabs.harvard.edu/abs/2007MNRAS.374..477T} {374, 477}

\bibitem[\protect\citeauthoryear{{Tinker}, {Kravtsov}, {Klypin}, {Abazajian},
  {Warren}, {Yepes}, {Gottl{\"o}ber}  \& {Holz}}{{Tinker}
  et~al.}{2008}]{2008ApJ...688..709T}
{Tinker} J.,  {Kravtsov} A.~V.,  {Klypin} A.,  {Abazajian} K.,  {Warren} M.,
  {Yepes} G.,  {Gottl{\"o}ber} S.,   {Holz} D.~E.,  2008, \mn@doi [\apj]
  {10.1086/591439}, \href
  {https://ui.adsabs.harvard.edu/abs/2008ApJ...688..709T} {688, 709}

\bibitem[\protect\citeauthoryear{{Tinker}, {Robertson}, {Kravtsov}, {Klypin},
  {Warren}, {Yepes}  \& {Gottl{\"o}ber}}{{Tinker}
  et~al.}{2010}]{2010ApJ...724..878T}
{Tinker} J.~L.,  {Robertson} B.~E.,  {Kravtsov} A.~V.,  {Klypin} A.,  {Warren}
  M.~S.,  {Yepes} G.,   {Gottl{\"o}ber} S.,  2010, \mn@doi [\apj]
  {10.1088/0004-637X/724/2/878}, \href
  {https://ui.adsabs.harvard.edu/abs/2010ApJ...724..878T} {724, 878}

\bibitem[\protect\citeauthoryear{{Tomooka}, {Rozo}, {Wagoner}, {Aung}, {Nagai}
  \& {Safonova}}{{Tomooka} et~al.}{2020}]{2020MNRAS.499.1291T}
{Tomooka} P.,  {Rozo} E.,  {Wagoner} E.~L.,  {Aung} H.,  {Nagai} D.,
  {Safonova} S.,  2020, \mn@doi [\mnras] {10.1093/mnras/staa2841}, \href
  {https://ui.adsabs.harvard.edu/abs/2020MNRAS.499.1291T} {499, 1291}

\bibitem[\protect\citeauthoryear{{Vikhlinin} et~al.,}{{Vikhlinin}
  et~al.}{2009}]{2009ApJ...692.1060V}
{Vikhlinin} A.,  et~al., 2009, \mn@doi [\apj] {10.1088/0004-637X/692/2/1060},
  \href {https://ui.adsabs.harvard.edu/abs/2009ApJ...692.1060V} {692, 1060}

\bibitem[\protect\citeauthoryear{{White}, {Cohn}  \& {Smit}}{{White}
  et~al.}{2010}]{2010MNRAS.408.1818W}
{White} M.,  {Cohn} J.~D.,   {Smit} R.,  2010, \mn@doi [\mnras]
  {10.1111/j.1365-2966.2010.17248.x}, \href
  {https://ui.adsabs.harvard.edu/abs/2010MNRAS.408.1818W} {408, 1818}

\bibitem[\protect\citeauthoryear{{Wilcox} et~al.,}{{Wilcox}
  et~al.}{2015}]{2015MNRAS.452.1171W}
{Wilcox} H.,  et~al., 2015, \mn@doi [\mnras] {10.1093/mnras/stv1366}, \href
  {https://ui.adsabs.harvard.edu/abs/2015MNRAS.452.1171W} {452, 1171}

\bibitem[\protect\citeauthoryear{{Wojtak} \& {{\L}okas}}{{Wojtak} \&
  {{\L}okas}}{2010}]{2010MNRAS.408.2442W}
{Wojtak} R.,  {{\L}okas} E.~L.,  2010, \mn@doi [\mnras]
  {10.1111/j.1365-2966.2010.17297.x}, \href
  {https://ui.adsabs.harvard.edu/abs/2010MNRAS.408.2442W} {408, 2442}

\bibitem[\protect\citeauthoryear{{Wojtak}, {{\L}okas}, {Mamon}  \&
  {Gottl{\"o}ber}}{{Wojtak} et~al.}{2009}]{2009MNRAS.399..812W}
{Wojtak} R.,  {{\L}okas} E.~L.,  {Mamon} G.~A.,   {Gottl{\"o}ber} S.,  2009,
  \mn@doi [\mnras] {10.1111/j.1365-2966.2009.15312.x}, \href
  {https://ui.adsabs.harvard.edu/abs/2009MNRAS.399..812W} {399, 812}

\bibitem[\protect\citeauthoryear{{Wojtak}, {Hansen}  \& {Hjorth}}{{Wojtak}
  et~al.}{2011}]{2011Natur.477..567W}
{Wojtak} R.,  {Hansen} S.~H.,   {Hjorth} J.,  2011, \mn@doi [\nat]
  {10.1038/nature10445}, \href
  {https://ui.adsabs.harvard.edu/abs/2011Natur.477..567W} {477, 567}

\bibitem[\protect\citeauthoryear{{Zhao}, {Peacock}  \& {Li}}{{Zhao}
  et~al.}{2013}]{2013PhRvD..88d3013Z}
{Zhao} H.,  {Peacock} J.~A.,   {Li} B.,  2013, \mn@doi [\prd]
  {10.1103/PhysRevD.88.043013}, \href
  {https://ui.adsabs.harvard.edu/abs/2013PhRvD..88d3013Z} {88, 043013}

\bibitem[\protect\citeauthoryear{{Zu} \& {Weinberg}}{{Zu} \&
  {Weinberg}}{2013}]{2013MNRAS.431.3319Z}
{Zu} Y.,  {Weinberg} D.~H.,  2013, \mn@doi [\mnras] {10.1093/mnras/stt411},
  \href {https://ui.adsabs.harvard.edu/abs/2013MNRAS.431.3319Z} {431, 3319}

\bibitem[\protect\citeauthoryear{{Zu}, {Weinberg}, {Jennings}, {Li}  \&
  {Wyman}}{{Zu} et~al.}{2014}]{2014MNRAS.445.1885Z}
{Zu} Y.,  {Weinberg} D.~H.,  {Jennings} E.,  {Li} B.,   {Wyman} M.,  2014,
  \mn@doi [\mnras] {10.1093/mnras/stu1739}, \href
  {https://ui.adsabs.harvard.edu/abs/2014MNRAS.445.1885Z} {445, 1885}

\bibitem[\protect\citeauthoryear{{Zwicky}}{{Zwicky}}{1937}]{1937ApJ....86..217Z}
{Zwicky} F.,  1937, \mn@doi [\apj] {10.1086/143864}, \href
  {https://ui.adsabs.harvard.edu/abs/1937ApJ....86..217Z} {86, 217}

\bibitem[\protect\citeauthoryear{{de Haan} et~al.,}{{de Haan}
  et~al.}{2016}]{2016ApJ...832...95D}
{de Haan} T.,  et~al., 2016, \mn@doi [\apj] {10.3847/0004-637X/832/1/95}, \href
  {https://ui.adsabs.harvard.edu/abs/2016ApJ...832...95D} {832, 95}

\bibitem[\protect\citeauthoryear{{van den Bosch}, {Norberg}, {Mo}  \&
  {Yang}}{{van den Bosch} et~al.}{2004}]{2004MNRAS.352.1302V}
{van den Bosch} F.~C.,  {Norberg} P.,  {Mo} H.~J.,   {Yang} X.,  2004, \mn@doi
  [\mnras] {10.1111/j.1365-2966.2004.08021.x}, \href
  {https://ui.adsabs.harvard.edu/abs/2004MNRAS.352.1302V} {352, 1302}

\bibitem[\protect\citeauthoryear{{van den Bosch}, {Lange}  \& {Zentner}}{{van
  den Bosch} et~al.}{2019}]{2019MNRAS.488.4984V}
{van den Bosch} F.~C.,  {Lange} J.~U.,   {Zentner} A.~R.,  2019, \mn@doi
  [\mnras] {10.1093/mnras/stz2017}, \href
  {https://ui.adsabs.harvard.edu/abs/2019MNRAS.488.4984V} {488, 4984}

\makeatother
\end{thebibliography}

\end{document}